\documentclass[AMA,Times1COL]{WileyNJDv5} 
\articletype{Article Type}%

\usepackage{multirow}
\usepackage{booktabs}
\usepackage{listings}
\usepackage{courier}
\usepackage{subcaption}
\usepackage{tikz}
\usetikzlibrary{positioning}
\usepackage{enumitem}
\usepackage{bm}
\usepackage{cleveref}
\AddToHook{cmd/appendix/before}{\crefalias{section}{appendix}}
\AddToHook{cmd/appendix/before}{\crefalias{subsection}{appendix}}
\AddToHook{cmd/appendix/before}{\crefalias{subsubsection}{appendix}}
\usepackage{makecell}

\usepackage{makecell}
\usepackage{siunitx}
\usepackage[svgnames]{xcolor}
\newcommand{\best}{\textsuperscript{$\dagger$}}
\newcommand{\approxdata}{\textsuperscript{$\ast$}}

\newcommand{\union}{\cup}

\newcommand{\prodarrow}{\;\;\rightarrow\;\;}
\newcommand{\rsep}{\;\;|\;\;}

\let\oldnl\nl
\newcommand{\nonl}{\renewcommand{\nl}{\let\nl\oldnl}}

\newcommand{\timeimprovement}{\text{8.15x}}
\newcommand{\memimprovement}{\text{18.10x}}

\definecolor{bgcolor}{HTML}{EFEFEF}

\newcommand{\mdeurl}{\url{https://aghorui.github.io/mde/}}

\received{XX XX XXXX}
\revised{XX XX XXXX}
\accepted{XX XX XXXX}
\journal{XXXX}
\volume{XX}
\copyyear{XXXX}
\startpage{1}

\begin{document}

\title{Points-to Analysis Using MDE: A Multi-level Deduplication Engine for Repetitive Data and Operations}

\author[1]{Anamitra Ghorui}

\author[2]{Aditi Raste}

\author[1]{Uday P. Khedker}

\authormark{GHORUI \textsc{et al.}}
\titlemark{Points-to Analysis Using MDE: A Multi-level Deduplication Engine for Repetitive Data and Operations}

\address[1]{\orgdiv{Department of Computer Science and Engineering}, \orgname{Indian Institute of Technology, Bombay},
\orgaddress{\state{Maharashtra}, \country{India}}}

\address[2]{\orgdiv{Department of Technology}, \orgname{Savitribai Phule Pune University},
\orgaddress{\state{Maharashtra}, \country{India}}}

\corres{
Anamitra Ghorui \email{aghorui@cse.iitb.ac.in}, \\
Aditi Raste \email{aditi.raste@campstud.unipune.ac.in}, \\
Uday P. Khedker \email{uday@cse.iitb.ac.in}}



\abstract[Abstract]{
Precise pointer analysis is a foundational component of many client analyses and optimizations. Scaling flow- and context-sensitive pointer analysis has been a long-standing challenge, suffering from combinatorial growth in both memory usage and runtime. Existing approaches address this primarily by reducing the amount of information tracked often, at the cost of precision and soundness. In our experience a significant proportion of this cost comes from propagation of duplicate data and low-level data structure operations being repeated a large number of times. Our measurements on SPEC benchmarks show that more than 90\% of all set-union operations performed can be redundant.

We present Multi-level Deduplication Engine (MDE), a mechanism that recursively augments the representation of data through de-duplication and the assignment of unique identifiers to values to eliminate redundancy. This allows MDE to trivialize many operations, and memoize operations enabling their future reuse. MDE's recursive structure allows it to represent de-duplicated values that themselves are constructed from other de-deuplicated values, capturing structural redundancy not easily possible with non-recursive techniques.

We provide a full C++ implementation of MDE as a library and integrate it into an existing implementation of a flow- and context-sensitive pointer analysis. Evaluation on selected SPEC benchmarks shows a reduction up to \memimprovement{} in peak memory usage and \timeimprovement{} in runtime. More notably, MDE exhibits an upward trend of effectiveness with the increase in benchmark size.

Besides performance improvements, this work highlights the importance of representation design and suggests new opportunities for bringing efficiency to future analyses.
}

\keywords{deduplication, memoization, data structures, program analysis, data-flow analysis, pointer analysis}

\jnlcitation{\cname{%
\author{Anamitra G.},
\author{Aditi R.},
\author{Uday P. K.}
}.
\ctitle{A Mechanism for Effective, Multi-level Memoization of Repetitive Data and Operations.} \cjournal{\it Software: Practice and Experience} \cvol{PREPRINT}.}

\maketitle

\renewcommand\thefootnote{}
\footnotetext{\textbf{Abbreviations:} BDD, binary decision diagrams; ZDD, zero-suppressed decision diagram; STL, standard template library}

\renewcommand{\thefootnote}{\arabic{footnote}}
\setcounter{footnote}{1}

\section{Introduction}
\label{sec:introduction}
In compiler design, automated static analyses of programs form a foundational element of program optimization. One of these analyses, pointer analysis, determines which locations a pointer may reference at a given point within the program. Results from a sound pointer analysis enable compiler transformation passes like dead-code elimination, loop invariant code motion, and global value numbering~\cite{web-llvmaainfra}. Pointer analysis results that are unsound are also useful~\cite{soundiness}, particularly in the detection of memory errors like use-after-free and null pointer dereferences. They are also used in client analyses such as taint analysis. However, the lack of soundness prevents their use in program transformation, where program semantics need to be preserved.

Historically, scaling interprocedural pointer analysis without compromising precision has been a challenging task. Since information accumulates combinatorially in a pointer analysis, a naive algorithm scales poorly in both memory usage and time. This is especially the case when the analysis takes into consideration the control-flow of a function (flow-sensitivity), and the calling contexts of functions (context-sensitivity). The precision of a pointer analysis is important because it allows its client analyses and transformations to make more informed and consequently, accurate decisions.

Compromising on the precision of the algorithm is a common approach to improving its scalability. For example, by making the analysis flow- or context-insensitive. This yields an imprecise, over-approximated summary of program behavior. Some implementations may also sacrifice soundness in favor of scalability. Despite there being a significant number of attempts at providing a scalable analysis in the past, achieving scalability while retaining both flow- and context-sensitivity remains a challenge. In particular, existing approaches primarily reduce the cost of analysis by limiting the information tracked, rather than eliminating redundancy present within that information.

\subsection{Past Approaches of Scaling Pointer Analysis}
\label{subsec:introduction:past-methods}
In past work, attempts at scaling program analysis have involved the following broad (not necessarily exhaustive) approaches:

\begin{enumerate}[label=A\arabic*:, ref=A\arabic*, font=\bfseries]
    \item \label{req:approx} Overapproximation of program information as done by flow-, context-insensitive algorithms and other methods.~\cite{andersen1994program, steensgaard1996points, bddbddb, graspan}

    \item \label{req:sparse} Making analysis data propagation sparse by the use of def-use chains, SSA-like techniques, and the use of cheap preanalyses.~\cite{sfspaper, svf-paper} \label{itemdefuse}

    \item \label{req:omitdata} Successively omitting irrelevant data-flow information as done by works like LFCPA~\cite{lfcpa} and VASCO.~\cite{vasco}

    \item \label{req:omitctl} Sound removal or summarization of redundant control flow as done by GPG-based points-to analysis.~\cite{gpg-analysis}

    \item \label{req:model} Performing analysis over individual elements of sets, instead of the entire set.~\cite{graspan, idfs-paper, ide-paper} This allows modeling of the problem as reachability over graphs, in which elements of the set form nodes.

    \item \label{req:datarepr} Removing redundancy in data and operations performed by changing the data structure used for representing analysis information.~\cite{bddbddb, semisparse-pta-paper, zddpta, barbar2021hash}
\end{enumerate}

These works, while beneficial, have not explored the possible benefits of a more comprehensive, concrete (as opposed to symbolic) encoding of pointer relationships.

In this paper, we propose another alternative to the above methods. This work is motivated by a set of empirical observations made by us and other works~\cite{svf-paper, barbar2021hash} in regards to the repetitive nature of the data and operations within an analysis explained in~\Cref{subsec:mdedef:motiv} and~\Cref{sec:related-work}, and the nature of the analyses presented by VASCO~\cite{vasco} and LFCPA~\cite{lfcpa}. Our past observations have found that as much as 72\% of all non-constant-time operations performed in the analysis (such as set unions, differences, set equality checking etc.) require using set equality checking operations (see~\Cref{tab:lfcpa-comparison}). Experimental results have shown that in many cases, more than 90\% of all union operations performed in an analysis are redundant (see~\Cref{subsec:gen-mde:additional-metrics}).

The main idea we present is that of a \textit{multi-level} or \textit{nested encoding} through the use of recursive de-duplication and unique identifiers as the basis for building the data structure used for the representation of data-flow values. Unlike previous approaches such as hash-consing,~\cite{barbar2021hash} our approach allows de-duplicated data items to themselves be constructed from other de-duplicated items, enabling a recursive structure in which sharing occurs across multiple levels of representation. This multi-level structure enables reuse of entire data-flow value instances, and reuse even when equality between data-flow values as a whole is absent, capturing structural redundancy that single-level approaches cannot exploit.

Experimental results of integrating the implementation of idea in our interprocedural flow- and context-sensitive pointer analysis implementation and evaluating on selected SPEC benchmarks shows a reduction up to \memimprovement{} in peak memory usage and \timeimprovement{} in runtime. More notably, MDE exhibits an upward trend of effectiveness with the increase in benchmark size.

\begin{figure*}[t]
\centering
\begin{subfigure}[t]{0.45\linewidth}
	\centering
	\includegraphics[width=0.8\linewidth]{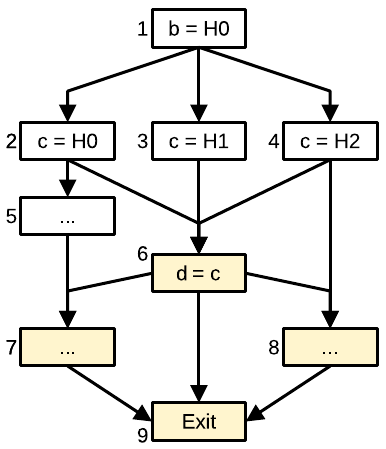}
	\caption{Control-flow graph with pointer assignments.}
	\label{fig:motiv_ctl}
\end{subfigure}
\hfill
\begin{subfigure}[t]{0.45\linewidth}
	\centering
	\includegraphics{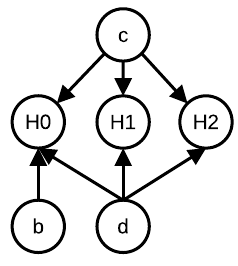}
	\caption{Pointer information in program points highlighted in yellow in~\Cref{fig:motiv_ctl}.}
	\label{fig:motiv_data}
\end{subfigure}

\caption{Motivating Example for Multi-level Deduplication.}
\label{fig:motiv}
\end{figure*}

\paragraph*{Motivating Example}
Consider~\Cref{fig:motiv_ctl} which depicts a control flow graph of a function. It contains pointer variables $b$, $c$ and $d$, and heap locations named $\text{H0}$, $\text{H1}$ and $\text{H2}$. If we were to run flow-sensitive pointer analysis on it, several of the program points (6, 7, 8, 9) would end up with an instance of the pointer relationship graph shown in~\Cref{fig:motiv_data}. If we run an interprocedural context-sensitive analysis as well, there is also a chance that the same pointer relationship graph may come up multiple times for the same nodes across calling contexts.

In the graph shown in~\Cref{fig:motiv_data},  the nodes $c$ and $d$ reference the same set of heap locations, or \textit{pointees}. If we were to represent this as an adjacency list, each program point stores its own copy of the nodes and edges (as a set of destination-nodes) in the graph. We may illustrate this information as:
\begin{equation}
 \{
 	c \rightarrow \{\text{H0},\text{H1},\text{H2}\},
 	d \rightarrow \{\text{H0},\text{H1},\text{H2}\},
 	b \rightarrow \{\text{H0}\}
 \} \label{eqn:level1}
\end{equation}
Assuming that each edge or node occupies one unit of memory, it takes 13 units per graph, and 52 units across the four program points. On a context-sensitive analysis, this number will be roughly $52 \times n$, where $n$ is the number of contexts.

A first level of optimization to reduce space is to de-duplicate identical sets of pointees. Since $c$ and $d$ share exactly the same pointees, their pointee sets can be stored once and referenced multiple times. If the reference to the set $\{\text{H0},\text{H1},\text{H2}\}$ is $\textit{Set1}$, and $\{\text{H0}\}$  is $\textit{Set2}$, then this representation becomes:
\begin{equation}
 \{
 	c \rightarrow \textit{Set1},
 	d \rightarrow \textit{Set1},
 	b \rightarrow \textit{Set2}
 \} \label{eqn:level2}
\end{equation}
Thus an instance of the graph now occupies the space of 3 nodes and 3 references. If a reference occupies one unit of memory, each program point would now hold 6 units of memory, totaling to 24 for the function, and roughly $24 \times n$ across all contexts. The cost we pay for it is a level of indirection to access the data.

Assuming that the data-flow values at a program point across different contexts are the same, this still stores multiple copies of structurally identical data-flow values at a program point across different contexts. By further de-duplicating the entire representation and replacing each instance with a reference to a shared resource in memory, we end up with each program point now storing only a single reference per context, reducing to 4 units in total, and roughly $4 \times n$ across all contexts. However, this representation adds another layer of indirection to access the data.

Note how this is a \textit{thirteen-fold} reduction in memory from the first approach and a \textit{six-fold} reduction from the second. If our observations about redundancy in analysis data are correct, it opens us to the possibility of being able to leverage this multi-level indirection despite the overheads that occur in this scheme. Being able to successfully exploit such multi-level redundancy can significantly reduce both memory usage and the cost of repeated operations.

\subsection{Multi-level Deduplication Engine}
\label{subsec:introduction:mde}
We present a Multi-level Deduplication Engine (MDE), which is a mechanism that can be applied to an existing data-flow analysis to memoize both data and operations performed in that analysis. MDE enables a recursive augmentation the representation of data through de-duplication and the assignment of unique identifiers to values to eliminate redundancy. This allows MDE to trivialize many operations, such as determining equality, and memoize other operations, enabling their future reuse. MDE's recursive structure allows it to represent de-duplicated values that themselves are constructed from other de-deuplicated values, capturing structural redundancy not easily possible with non-recursive techniques. By exploiting such partial and structural redundancy, MDE enables reuse at multiple levels of a given representation.

In~\Cref{sec:insights}, we provide informal reasoning and potentially important insights as to why a multi-level memoization system like this, and memoization in general, may strictly consume less memory and processing time compared to a default implementation using C++ Standard Template Library (STL), given that certain conditions are satisfied.

We believe Approach~\ref{req:omitdata} and Approach~\ref{req:omitctl} benefit greatly from a system as described by our work. LFCPA and VASCO especially involve the liberal use of data-flow values, along with the storage of the data-flow values themselves across the analysis in order to achieve scalability. It therefore makes sense to have a system to de-duplicate such data and operations on such data. We provide additional commentary and rationale for this, and other works that have used custom data structures~\Cref{sec:related-work}.

\paragraph*{Generalization.} While we consider it to be outside the scope of this paper, we have extended this system for potential general usage for any form of activity that is similar to that found in compiler design or program optimization. We detail this generalized system and several potential uses starting from~\Cref{sec:gen-mde}.

\subsection{Main Contributions}
We make the following contributions in this work:
\begin{enumerate}
\item A novel dimension of de-duplication based on representation of data in the context of program analysis.
\item A multi-level memoization technique to an interprocedural, flow- and context-sensitive pointer analysis, enabling recursive de-duplication of data-flow values.
\item A full-featured C++ library implementing the approach for pointer analysis and for potential applications beyond pointer analysis.
\item An empirical evaluation of multi-level deduplication for flow- and context-sensitive pointer analysis.
\end{enumerate}

\section{Background on Data-Flow Analysis and Pointer Analysis}
\label{sec:background}
This section explains several concepts and prior work on which MDE stands, and are important to understanding MDE and its area of application.

\paragraph*{Data-flow Analysis.} Data-flow analysis is a type of program analysis wherein we observe how certain facts about the program's variables and memory locations propagate through its control flow. Commonly, this is done through an iterative computation across all points of the program, and repeated until the analysis reaches a fixed point, or when there is no longer any change in the data.

\paragraph*{Live Variables Analysis.}
\textit{Live Variables Analysis}, sometimes referred to as Liveness Analysis, attempts to determine whether a variable will be potentially used in the future or not starting from a certain program point, that is, whether the variable is live or not. We represent the live variables at a given point as a set. For example, if the variables $a$, $b$, and $c$ are live at a certain point in the program, we represent this fact, or \textit{data-flow value}, as:
\[
\{ a, b, c \}
\]
To mark a variable as live or otherwise, we consider statements of the program for their semantic meaning. For example, if there is a statement like $a = b$ in the program, and there are no other live variables after the statement, we say that $b$ is live in all paths before this statement (since its value is being assigned), and $a$ is no longer live before this statement (since its value is being overwritten). Thus the final set of live variables \textit{before} this statement (the `IN' set) is $\{b\}$, and \textit{after} this statement (the `OUT' set) is $\emptyset$. If there were live variables after this statement, we represent them by copying them into `OUT', and then process the semantic meaning of the statement using that information to compute `IN'. For example, if $a$ was present in `OUT' (i.e., is live after the statement), it would not be present in `IN', since $a$ is being assigned to in the statement. We repeat this computation across all statements until the `IN' and `OUT' sets of all of them no longer change, that is, when the computation reaches a fixed point.

As can be possibly inferred from this description, information flows \textit{backward} in this analysis. While computation will eventually reach a fixed point in whatever order the statements are processed. it is more efficient to perform this fixed-point computation in a direction opposite to the direction of the control flow.

\paragraph*{Pointer Analysis.}
\textit{Pointer Analysis} or points-to analysis, on the other hand, is a \textit{forward} analysis, where we compute the `IN' set first, and the `OUT' set second. Instead of representing variables that are live, we compute where does a pointer variable point to. Pointer analysis forms a foundation for many client optimization passes where it is important to determine the uniqueness and usages of heap locations, such as dead store elimination, dead code elimination, global value numbering, and so on~\cite{web-llvmaainfra}.

We will use two abstract representations of this information: \textit{points-to pairs} and \textit{points-to map}. For example, at a particular point in the program, if the variable $a$ may point to $b$ and $c$, and $c$ may point to $d$, the points-to pair representation of this information would look like:
\[
\{ a \rightarrow b, a \rightarrow c, c \rightarrow d \}
\]
The points-to map, which is essentially an adjacency list representation of the relationships, instead would look like:
\[
\{ a \rightarrow \{ b, c \}, c \rightarrow \{ d \} \}
\]
The set of \textit{pointees}, which are $\{b, c\}$ and $\{d\}$ in this case, are commonly referred to as a \textit{points-to set}. To avoid confusion, we shall only refer to this as the set of pointees, or the pointee-set. We discuss the significance of these two representations in later sections.

\subsection{Scalability of Pointer Analysis}

Due to the nature of how pointer relationships are made, the amount of information can grow non-linearly, or even combinatorially, with the number of statements in the program. \Cref{tab:lfcpa-comparison}~shows experimental results using our analysis listing the number of operations performed against benchmarks in the SPEC suite~\cite{speccpu2006}. As can be seen, even for small benchmarks, the number of operations blows up significantly with program size. Being able to scale pointer analysis is therefore a challenge. Earlier works\cite{andersen1994program, steensgaard1996points, bddbddb} also highlight this problem. Note also, how the table particularly indicates an overwhelming contribution of equality comparisons in the count. As much as 72\% of the operations in the \texttt{hmmer} benchmark are just equality comparisons. The high percentage of equality comparisons is one of the attributes exploited by our technique.

\begin{table*}[t]
    \centering
    \sisetup{scientific-notation = true}
    \begin{tabular}{lrrrrrr}
        \toprule
        \textbf{Name} &
        \textbf{\makecell{IR Size\\(Bytes)}} &
        \textbf{\makecell{IR Line\\Count}} &
        \textbf{\makecell{Unique\\Pointee Sets}} &
        \textbf{\makecell{Unique\\Points-to Maps}} &
        \textbf{Operation Count\textsuperscript{*}} &
        \textbf{Equality Comparison Count}  \\
        \midrule
        \texttt{lbm}
        	& 428K  
        	& \num{7155} 
        	& \num{4255} 
        	& \num{1124} 
        	& \num{117108} 
        	& \num{44948} (38.4\%) \\ 
		\texttt{mcf}
			& 512K 
			& \num{7483} 
			& \num{17458} 
			& \num{53016} 
			& \num{2788321} 
			& \num{391485} (14.0\%)  \\ 
		\texttt{libquantum}
			& 1.1M 
			& \num{16332} 
			& \num{6086} 
			& \num{1486} 
			& \num{1224761} 
			& \num{490527} (40.1\%)   \\ 
		\texttt{bzip2}
			& 2.4M 
			& \num{36591} 
			& \num{89971} 
			& \num{638396} 
			& \num{70605717} 
			& \num{15082646} (21.4\%)    \\ 
		\texttt{sjeng}
			& 1.7M 
			& \num{39021} 
			& \num{37657} 
			& \num{272902} 
			& \num{313804855} 
			& \num{130348713} (41.5\%)   \\ 
		\texttt{hmmer}
			& 9.5M 
			& \num{150611} 
			& \num{491436} 
			& \num{1073174} 
			& \num{413184235} 
			& \num{298455856} (72.2\%) \\ 
        \bottomrule
    \end{tabular}
    \begin{tablenotes}
    \textsuperscript{*}"Operations" here refer to the non-constant-time operations performed by the analysis on data-flow values. This includes performing set unions, differences, comparing sets, determining set equality and criteria-based filtering of set elements.
    \end{tablenotes}
    \sisetup{scientific-notation = false}
    \caption{Experimental results of running the VASCO-LFCPA analysis implementation on selected SPEC benchmark programs. Results are provided for the MDE-based implementation (see~\Cref{sec:evaluation}). File statistics are provided for the generated LLVM intermediate representation (IR) file rather than the source. The percentage of operations that are equality comparisons are provided in parantheses in the last column. Data is arithmetic mean of 7 observations.}
    \label{tab:lfcpa-comparison}
\end{table*}

\paragraph*{Flow- and Context-Sensitivity.}
In order to enable a scalable pointer analysis, several early established works on it~\cite{andersen1994program, steensgaard1996points} have made compromises in regards to the precision of the analysis by removing consideration of the control flow of the program (flow-sensitivity), as well as the context of the calling sites of a function (context-sensitivity). While this results in a comparatively faster analysis, the imprecision of the result makes it not very useful for performing optimizations.

Another point to note is that analyses may forego \textit{soundness} of the result entirely to achieve scalability~\cite{soundiness}. This keeps the results useful for memory error detection such as null pointer dereferences and use-after free, but prevent their use in optimization or transformation since program semantics need to be preserved. Our pointer analysis implementation attempts to be sound and precise, while still being scalable.

\subsection{LFCPA, VASCO and VASCO-LFCPA}
\label{sec:background:pta}
Liveness-Based Flow- and Context-Sensitive Points-to Analysis, or LFCPA~\cite{lfcpa} is an augmentation of normal flow- and context-sensitive pointer analysis and is what we use in our C++ pointer analysis implementation. To enable interprocedural analysis, we further augment LFCPA with \textit{Value-Sensitive Contexts}, or VASCO~\cite{vasco}. Henceforth, we will refer to the combined mechanism as \textit{VASCO-LFCPA}. Here, we only provide brief explanations that are relevant to MDE.
\paragraph*{General Idea of LFCPA.}
We use, in alternation, liveness analysis results of pointers in points-to analysis and vice versa to determine whether we should actually compute points-to information or not, cutting down the information we actually need to propagate. Thus, instead of simply propagating points to information across the control flow, we also propagate variable liveness information. A sample representation of data flow information at a program point may look like the following pair of two sets. The first set of the pair represents points-to information, whereas the second represents liveness information.
\[
(\lbrace
	p1 \rightarrow \lbrace a, b \rbrace
\rbrace, \lbrace p1, p3 \rbrace)
\]
\paragraph*{General Idea of VASCO.}
To enable interprocedural analysis and context-sensitivity, one of the classical methods is to differentiate contexts with call strings,~\cite[Chapter~7]{callstrings} which essentially captures the chain of unfinished calls that have been made up to the entry point of a function.\footnote{There are special considerations made in the case of recursion.} Call strings can turn out to be large and numerous, especially in large programs. Practical implementations often limit the size of the call strings by using only the latest N elements of it, decreasing the precision of the contexts. As an alternative, we may use value-sensitive contexts,~\cite{vasco} where we instead capture the incoming data flow values at a given callsite and differentiate calling contexts on that basis. In the case of LFCPA, these would be the incoming points-to and liveness information.

A point to note is that an interprocedural, context-sensitive analysis has to be \textit{re-run} for a procedure, in each distinct context of that procedure, and therefore, a separate set of data-flow values is computed and stored for each context, contributing to information blow-up.

\section{Description of MDE}
\label{sec:mdedef}

Multi-level Deduplication Engine (MDE) is a deduplication mechanism that transparently replaces operations, as well as data that flows through a data-flow analysis through minimal refactoring of existing code. Deduplication of data and memoization of operations is done in hopes that we gain both time and space efficiency in the analysis, by preventing existence of repeated data and calculations. The current implementation of MDE is in C++.~\footnote{\mdeurl} For our specific implementation in points-to analysis, our aim is to do the following:
\begin{enumerate}
\item Make calculation of liveness results more efficient,
\item Make calculation of pointee-set results more efficient, and
\item Make calculation of points-to map results as a whole more efficient.
\end{enumerate}
We seek efficiency in both memory usage and time in all of these cases.

\subsection{Motivations}
\label{subsec:mdedef:motiv}
Despite the augmentations made for VASCO-LFCPA being a net improvement over standard flow and context-sensitive pointer analysis, the issue of scalability in large programs persists. While attempts at improvement are still being made from the algorithmic perspective in our implementation, other avenues of improvement have been overlooked. It remained to be seen whether the lack of scalability of VASCO-LFCPA is due to the algorithm itself or the implementation.

C++ STL data structures (\texttt{std::map} and \texttt{std::set}) were used to store data-flow values. Points-to information in particular was stored in a points-to map representation as a \texttt{std::map} of variables to \texttt{std::set}s. Because of the nature of these data structures, time complexities upto $O(n)$ (where $n$ is the number of data-flow value elements) were incurred in the implementation when performing common operations, such as checking set equality, performing unions and intersections, and so on. In many places, ad-hoc $O(n)$ operation implementations were being used where $O(1)$ would suffice. Since the number of elements increases combinatorially, performing these operations became a major bottleneck.

This problem motivated a refactoring of the existing implementation to use better data structures or mechanisms. Ideas for it were based on three empirical observations about analysis computations:
\begin{enumerate}
\item Data-flow information is usually very sparse in programs. This is evidenced by the high number of ``empty hits'' shown in~\Cref{tab:sample-statistics}.
\item A large number of duplicates exists in information across the control flow as a consequence, as evidenced by a high number of ``equal hits'' shown in~\Cref{tab:sample-statistics}.
\item Much of the computations performed in an analysis are redundantly performed on the same data, as shown by the overall ``hit ratio'' in~\Cref{tab:sample-statistics}.
\end{enumerate}

\subsection{First Idea of the Mechanism}
\label{subsec:background:first-idea}
One method of exploiting the above observations for gaining efficiency in an analysis could be to perform a straightforward de-duplication of the data flowing through it:
\begin{enumerate}
\item Assign a unique integer to each operand/variable, and use a pair of these integers as a points-to pair.
\item Assign a unique integer to each points-to pair, and put these integers into a sorted vector to use as our points-to information. Use the integer identifiers for individual operands in a sorted vector as our live variable set.
\item Assign unique integers to the sets of points-to pairs and live variable sets, and use them as our points-to and liveness representation in the analysis respectively.
\end{enumerate}
This approach allows us several advantages despite introducing additional levels of indirection for access. The first being that any attribute-based comparisons of individual operands are ruled out. The second being that comparisons between points-to maps are reduced to an integer comparison. The third is that, by virtue of being a sorted list, the majority of aggregate operations on the sets are $O(n)$ in the worst case. The deduplication mechanism used to perform this activity is described in~\Cref{subsec:mdedef:dedup}. 

\paragraph*{Shortcomings.} However, this approach suffers from several limitations. While this approach lets us reduce the effective amount of redundancy in the system, it still does not capture the redundancy that may occur within the set of pointees in each pointer. It also was not clear whether operations like replacing the entire set of pointees for a given pointer would be a bottleneck with this representation. Another big problem is that the points-to pair representation is not compatible with the points-to map representation that VASCO-LFCPA uses, and introduction of it would have essentially amounted to a full rewrite. This led us to devising a new approach which is the one defined by MDE.

\subsection{Defintion and Construction of MDE}

This section describes the basic structure of MDE. We first start by defining all of the foundational components of MDE, and then describe how the recursive, or nested encoding of points-to map is constructed in~\Cref{sec:nestedcons}.

Deriving from the nature of the data in these analyses, the mechanism operates on two elementary units: the \textit{property}, and the \textit{property set}. A property is a user-defined type. In the case of liveness analysis, this may be a an object representing a variable. A property set is a collection of unique properties. Property sets are the unit of operations in MDE, and are supplied as operands. All property sets that have been registered within MDE are treated as \textit{immutable}. The main reason for this is that it allows us to store property sets as \textit{sorted vectors}, on which many common set operations can be performed in $O(n)$ amount of time. The second reason is that the same set may be used unchanged at some other point within the program being analyzed.

\subsubsection{The Deduplication Sub-Mechanism}
\label{subsec:mdedef:dedup}

This mechanism is used in MDE to deduplicate property sets. The main purpose of the deduplicator is to assign a unique identifier to each unique data item given to it. We call this process a \textit{registration}. It has two main sub-components: a storage vector and a hash table performing the mapping between items and IDs. We interchangably refer to the unique identifier as an \textit{index}.

\paragraph*{Structure of the Mapping Table.} For it to uniquely identify a data item, the deduplicator mandates two routines from the user:

\begin{description}
\item[Equality Comparator:] For simple data types, this may be simple in nature and may not even require explicit specification. However, in large or aggregate data structures like the property sets in MDE, such comparators have to be explicitly implemented and specified. It also implies that this is not necessarily a constant-time operation.
\item[Hash Function:] The hash table requires this for $O(1)$ lookup of values. Like in the case of the equality comparator, this is also not necessarily an $O(1)$ operation.
\end{description}

This also exposes another problem: the storage of keys themselves. The data items, as alluded to, can get large in size, and storing them as the keys of the item-to-integer mapping may result in very slow behavior if the hash table is reallocated or moved around, or the key-value pairs themselves are relocated.

Another problem lies in the case of performing a lookup. If the hash table implementation passes copies of the lookup parameter to the hashing function or any other internal component, this will also increase the amount of memory and time overhead of the deduplicator implementation. Hence, we keep pointers as our keys instead, making it a mapping from a pointer to an integer. The places we make these pointers point to are the storage locations in memory held by the storage vector made when a new item is registered. The pointers, therefore, subsequently map to their respective identifier.

\paragraph*{Structure of the Storage Vector.} The storage location of the actual data item is made to never change for the lifetime of the deduplicator object. This is handled by the storage vector, which is a resizable vector of C++ STL \texttt{std::unique\_ptrs} (smart pointers) that will free the held object upon destruction. The reason we don't simply just directly store the items as-is, is because relocating or copying the data item may be expensive, the semantics of the data item may not allow relocation or copying (e.g, the object has no C++ copy or move constructors), or cause undefined behavior. For example, if we were to directly store pointers to elements of the vector in the mapping table, these pointers would be invalidated once the vector reallocates, causing them to become dangling pointers. This makes the smart pointer approach more sound and easier to use.

\paragraph*{Deduplication Flow Example.} Let's say the user wants to register some item $A$, and $A$ is currently not in the deduplicator. The mapping table takes a pointer to $A$, tries to look it up, and does not find it. The registration routine then proceeds to create a new \texttt{std::unique\_ptr} object, and copies or moves $A$ into it depending on what the user prescribed as a parameter. The \texttt{std::unique\_ptr} object is then inserted into the storage array, and a new mapping is inserted with the storage location held by \texttt{std::unique\_ptr} as the key, and its index of insertion as the value. Since this is a vector, the index acts as both an identifier and the location in the vector where the object can be found. Finally, the user is returned the index as the canonical identifier for $A$. If $A$ was already present in the deduplicator, the index found in the hash table will be returned.

\paragraph*{Efficiency.} While this incurs an $O(n)$ cost of hashing, equality lookup, and an $O(n)$ cost of allocating new space for property sets, we mitigate this by exploiting the repetitiveness of the data through the aggressive memoization that we perform on it. We depend on the data being sparse and redundantly computed and compared to obtain gains in efficiency. We provide more reasoning for the presence of efficiency in~\Cref{sec:insights}.

\paragraph*{Notation.} We will use the following pseudocode notation to denote activities in the deduplicator:

\begin{itemize}
\item $\texttt{MDEState}$ denotes the collection of all elementary data structures and members in a MDE instance. We provide a full definition of this later on.
\item $\texttt{Register(MDEState, PropSet)}$ registers a property set \texttt{PropSet} into the specified MDE if not already present, and returns the unique identifier for it.
\end{itemize}

In general, for maps, we shall use \texttt{Map.get(K)} to retrieve the corresponding value of key \texttt{K}, \texttt{Map.set(K, V)} to assign key \texttt{K} to \texttt{V} and \texttt{Map.has(K)} to check for existence of a mapping for \texttt{K}.

For vectors, we shall use \texttt{PushOne(Vec, Elem)} to denote the insertion of a single element to the end of the vector, and \texttt{PushRange(Vec, Begin, End)} to denote the insertion, to the end, the elements from a C++-style iterator specification.

\subsubsection{The Operation Memoization Sub-Mechanism}
\label{subsec:def:opermemo}
This is the second main mechanism used in MDE, and is used in conjunction with the deduplication mechanism described in the earlier section. MDE defines operations specifically on property sets, such as performing unions, intersections and set differences. The mechanism consists of two components:

\begin{enumerate}
\item A hash table that maps a tuple of property set indices, which represents the operands of the operation, to another index, representing the result of the operation.
\item An implementation of the operation on the property set, say, as a C++ function.
\end{enumerate}

When MDE is queried for the result of an operation, say $\texttt{F}$, with operands $(\texttt{P1}, \texttt{P2}, \texttt{P3}, \texttt{P4}, ...)$, we first perform a lookup on its corresponding map. If the result exists, we return the mapped result index. If it does not, we call $\texttt{F}$ to actually perform the operation, and then $\texttt{Register}$ the result within MDE. Commutativity, associativity, and other algebraic properties of an operation are handled individually by each implementation of that operation in MDE. We explain this further in \Cref{subsec:mdedef:opersemantics}.

\paragraph*{Notation.} We will use the following pseudocode notation for operations:

\begin{itemize}
\item $\texttt{Operate(MDEState, F, P1, P2, ...)}$ returns the index of the result of $\texttt{F(P1, P2, ...)}$.
\item $\texttt{F.Map}$ refers to the corresponding memoization map of $\texttt{F}$.
\end{itemize}

\subsubsection{Description of Components}
\begin{figure*}[t]
	\centering
	\includegraphics[width=0.8\linewidth]{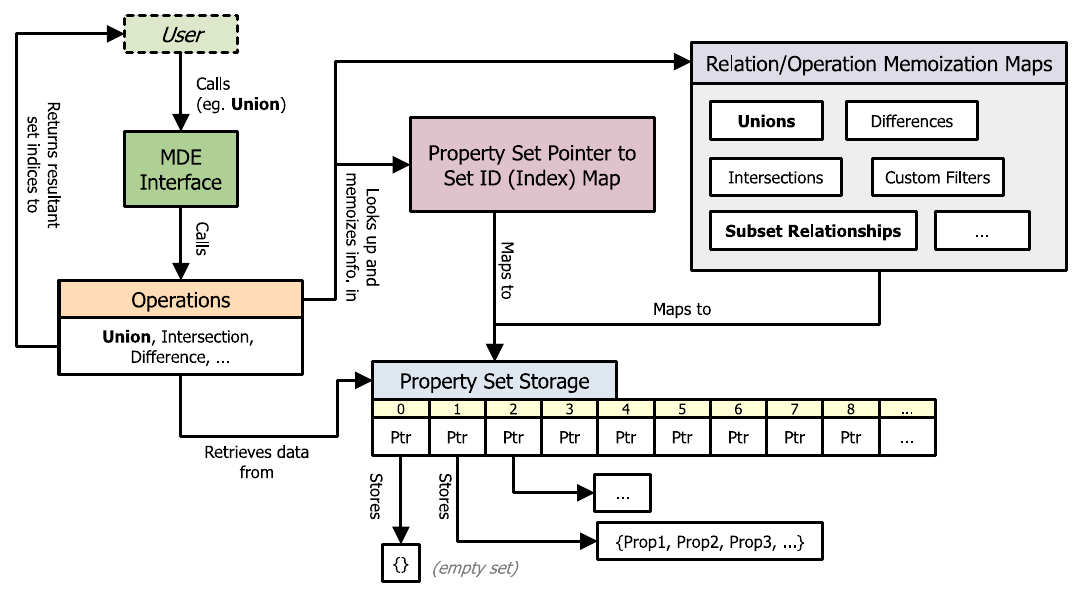}
	\caption{Schematic diagram of the state of an individual MDE. Subcomponents concerned with performing, as an example, the union operation, are highlighted in bold.}
	\label{fig:mde-schematic}
\end{figure*}

Based on the elementary descriptions of the deduplication and operation memoization sub-mechanisms, we now define the structure of the basic form of MDE. \texttt{MDEState} consists, firstly, of an instance of the deduplicator mechanism, and thus includes a hash table mapping property sets to indices, called \texttt{PropertySetMap}, and a storage array, called \texttt{Storage}. Secondly, it includes several set operations and their corresponding hash-maps. Finally, it includes a map called \texttt{SubsetRel}, that memoizes information about whether one set is the subset or the superset of the other. An illustration of the components is provided in~\Cref{fig:mdestate}, and as a schematic in~\Cref{fig:mde-schematic}. The initial state of the MDE contains a single registered property set with the index $0$, which is the \textit{empty set}.

The actual C++ implementation consists of a substantial amount of template metaprogramming to achieve static resolution of many components. For the sake of brevity, we provide a simplified view of them.

\begin{figure}[t]
	\begin{lstlisting}[numbers=none, xleftmargin=0ex]
MDEState<Property> {
	// Property Set Definition:
	using PropertySet = Vector<Property>;
	// Deduplicator objects:
	HashMap<PropertySet*, Index> PropertySetMap;
	Vector<UniquePointer<PropertySet>> Storage;
	// Operations and Maps:
	Union; Intersection; Difference; ... ;
	HashMap<(Index, Index), Bool> SubsetRel;
}
	\end{lstlisting}
	\caption{List of objects defining the state of MDE.}
	\label{fig:mdestate}
\end{figure}

\subsubsection{Description of Operation Semantics}
\label{subsec:mdedef:opersemantics}
The constraints that an implementation of an operation in MDE must maintain is that for a given tuple of operands, it should return a consistent, static index to a valid property set registered within MDE as a result. The means of achieving this is left up to the implementation. Here, we provide a description and algorithm of performing a two-operand union operation on property sets. The algorithm can be similarly modified for implementing other operations, such as intersections or differences. These algorithms are derived from C++ STL implementations on ordered containers (e.g. \texttt{std::set\_union}~\cite{web-setunioncpp}).

As mentioned earlier, property sets are represented as immutable sorted vectors. This allows us many operations in $O(n)$ amount of time in a similar fashion to the merge operation in merge sort~\cite{wiki-mergesort}. We illustrate this in~\Cref{algo:setunion}. In essence, we allocate a result vector, create two ordered iterators for both the operands, then iterate and compare values for equality, and the less-than relation. We append equal values only once in the result array, and if two values are unequal, we use the less-than relationship to determine the ordering. Thus, the resultant vector is also sorted. This, of course, implies that the type of the property operated on by MDE has to be \textit{totally ordered}.

There are however more considerations to make for efficiency outside of this algorithm and to exploit memoization and trivial deductions:

\paragraph*{Empty and Equal sets.}
We must check if one of the sets is empty, or if both of them are equal, before redundantly performing the operation as the result is trivially deducible. This is one of the reasons why we mark the empty set with the index 0.

\paragraph*{Subset relations.}
We must check if one of the sets is a subset or superset of the other as this may make the operation trivially deducible. We use the subset map to check this. If a tuple $(a, b)$ exists in the map, and is mapped to \texttt{true}, we say that $a$ is a subset of $b$. If mapped to \texttt{false}, we say that $a$ is a superset of $b$. If the tuple is not present at all, we say we don't know about this relationship, or that it may not hold at any time. To prevent duplicates of relations, the tuple is always stored such that the numerical value of $a$ is less than $b$.

This map is populated by the operations themselves. For example, if the union of $a$ and $b$ is $c$, we immediately store the relations $(a, c) \rightarrow \texttt{true}$ and $(a, b) \rightarrow \texttt{true}$ in the map.

\paragraph*{Redundant operations.} We must check if an operation has already been performed as detailed in~\Cref{subsec:def:opermemo}. This involves querying the corresponding map of the operation. If there is a hit, the mapped index is returned. If not, the actual operation is performed. If the operation is \textit{commutative}, like in the case of set union and intersection, we enforce a numerical ordering on the operand tuples like we have in the subset map. If the operation is not commutative, like in the case of set difference, we do not enforce ordering as it would be logically inconsistent.

\paragraph*{Final Algorithm.} The final algorithm for the union operation implementation in MDE is given in~\Cref{algo:fulloper}. Note that the procedure the algorithm is contained in is called \texttt{OperateUnion} for illustrative purposes. This can be similarly expressed in our notation as the call \texttt{Operate(State, Union, P1, P2)}.

\paragraph*{Note on Associativity.} The associativity of operations of more than two operands is left up to the implementation of the operation, as it would be unnecessarily complicated to do as a part of the internal API of MDE.

\paragraph*{Note on Closure.} By the nature of the structure of MDE, operations are closed under the domain of property sets of the user-specified property type. In a sense, it could be potentially said that MDE forms an algebraic structure. We elaborate on this in~\Cref{subsec:gen-mde:parallels}.

\paragraph*{Extending MDE to Other Operations.} The redundancy checks that we mention above are not necessarily applicable in all possible operations, or there may be valuable redundancy checks specific to a particular operation. Hence, enforcing minimal constraints on the operation implementation allows us to have the flexibility to do so. The implementer has the liberty to include more structures in MDE to facilitate exploitation of redundancy.

\begin{algorithm*}[t]
\caption{Set Union Algorithm}
\label{algo:setunion}
\begin{algorithmic}[1] 
\Procedure{Union}{$first, second$}
    \State \textbf{Input:} $first, second$
    \State \textbf{Output:} $new\_set$
    \State $new\_set \leftarrow \emptyset$
    \State $cursor_1 \leftarrow first.\text{begin}()$; $cursor\_end_1 \leftarrow first.\text{end}()$  \Comment{Iterators for first set}
    \State $cursor_2 \leftarrow second.\text{begin}()$; $cursor\_end_2 \leftarrow second.\text{end}()$ \Comment{Iterators for second set}

    \While{$cursor_1 \neq cursor\_end_1$}
        \If{$cursor_2 = cursor\_end_2$} \Comment{Push remaining data from first set}
            \State \Call{PushRange}{$new\_set, cursor_1, cursor\_end_1$}
            \State \textbf{break}
        \EndIf

        \If{$*cursor_2 < *cursor_1$}
            \State \Call{PushOne}{$new\_set, *cursor_2$}
            \State $++cursor_2$
        \Else
            \If{$\neg (*cursor_1 < *cursor_2)$}    \Comment{Both elements are equal at current position}
                \State $++cursor_2$
            \EndIf
            \State \Call{PushOne}{$new\_set, *cursor_1$}
            \State $++cursor_1$
        \EndIf
    \EndWhile

    \State \Call{PushRange}{$new\_set, cursor_2, cursor\_end_2$} \Comment{Push remaining data from second set}
    \State \Return $new\_set$
\EndProcedure
\end{algorithmic}
\end{algorithm*}

\begin{algorithm*}[t]
\caption{Full algorithm for Union with checks.}
\label{algo:fulloper}
\begin{algorithmic}[1] 
\Procedure{OperateUnion}{$State, P1_{\text{Orig}}, P2_{\text{Orig}}$}
    \State \textbf{Input:} $State, P1_{\text{Orig}}, P2_{\text{Orig}}$
    \State \textbf{Output:} Index

    \If{$P1_{\text{Orig}} = P2_{\text{Orig}}$} \Comment{Equality check}
        \State \Return $P1_{\text{Orig}}$
    \EndIf

    \State $P1 \leftarrow \min(P1_{\text{Orig}}, P2_{\text{Orig}})$; $P2 \leftarrow \max(P1_{\text{Orig}}, P2_{\text{Orig}})$ \Comment{Ordering check}

    \If{$P1 = 0$} \Comment{Empty set check}
        \State \Return $P2$
    \EndIf

    \If{$State.\text{SubsetMap.has}((P1, P2))$}
        \State $rel \leftarrow State.\text{SubsetMap.get}((P1, P2))$
        \If{$rel$} \Comment{$P1 \subset P2$}
            \State \Return $P2$
        \Else \Comment{$P1 \supset P2$}
            \State \Return $P1$
        \EndIf
    \EndIf

    \If{$State.\text{UnionMap.has}((P1, P2))$} \Comment{Redundancy check}
        \State \Return $State.\text{UnionMap.get}((P1, P2))$
    \EndIf

    \State $first \leftarrow State.\text{Storage.get}(P1)$; $second \leftarrow State.\text{Storage.get}(P2)$
    \State $new\_set \leftarrow State.\Call{Union}{first, second}$ \Comment{Actually perform operation}
    \State $new\_index \leftarrow \Call{Register}{State, new\_set}$ \Comment{New data registration}

    \State $State.\text{UnionMap.set}((P1, P2), new\_index)$
    \State $State.\text{SubsetMap.set}((P1, new\_index), \mathbf{true})$
    \State $State.\text{SubsetMap.set}((P2, new\_index), \mathbf{true})$

    \State \Return $new\_index$
\EndProcedure
\end{algorithmic}
\end{algorithm*}

\subsubsection{Miscellaneous API Routines}
This section lists some other routines present in the MDE API.

\begin{itemize}
\item \texttt{Contains(Idx, Prop)} returns whether \texttt{Prop} exists within the property set with index \texttt{Idx}. For sets of size below a certain threshold, a linear search is used. Above it, a binary search is used for finding \texttt{Prop}.
\item \texttt{InsertSingle(Idx, Prop)} returns a new index with \texttt{Prop} included into the property set with index \texttt{Idx}. This is a wrapper around performing a set union and is used to replicate common idioms in existing code that may be hard to replace. However, it may be more efficient for the client code to first accumulate data in an ordered container, and then pass an iterator to MDE for registering the set.
\item \texttt{RemoveSingle(Idx, Prop)} is similarly a wrapper around performing a set difference.
\end{itemize}
\subsubsection{Sample Demonstration}
\label{subsec:mdedef:nonnestdemo}
\begin{figure*}[t]
	\centering
	\includegraphics[width=0.8\linewidth]{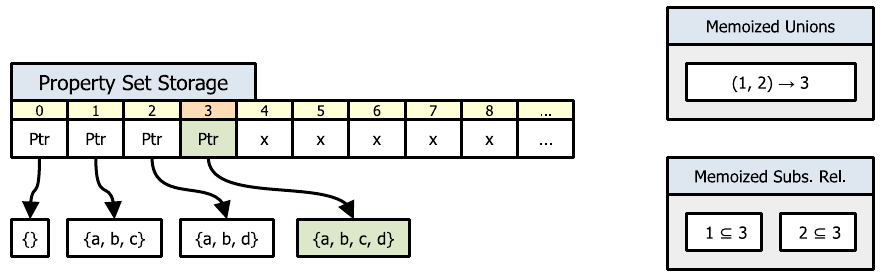}
	\caption{Final configuration of MDE after peforming the steps in~\Cref{subsec:mdedef:nonnestdemo}.}
	\label{fig:mde-nonnested-final-state}
\end{figure*}

Let us consider an empty \texttt{MDEState} that is configured to store and operate on sets of integers, and we want to perform the union of $\{a, b, c\}$ and $\{a, b, d\}$.
\begin{enumerate}
\item First, we \textit{register} $\{a, b, c\}$ and then $\{a, b, d\}$ into it. Assume that this gives us the corresponding indices $1$ and $2$, respectively, by the deduplication process in \texttt{PropertySetMap}.
\item Next, to actually perform the union, we perform a call to the union operation with these two identifiers.
	\begin{enumerate}
	\item First, MDE checks whether either of the operands is $0$, which is a special index reserved for the empty set.
	\item Seeing this isn't the case, MDE tries to look up the tuple $(1, 2)$ in the union operation map, but fails.
	\item Next, it looks in the subset map if the operation can be trivially deduced with a subset relation, but fails again.
	\item  It determines that it has to actually perform the operation, which it does and creates the new set $\{a, b, c, d\}$. This gets registered in the MDE with the index $3$, and this is the result that gets returned to the user.
	\end{enumerate}
\item  MDE records this new union that has been performed, as well as the subset relations inferred from the operation. Any future requests for the same or adjacent operations will result in the memoized index being returned. Namely, it records that the union of indices $1$ and $2$ to be $3$, and that $3$ is the superset of $1$ and $2$.
\end{enumerate}
We depict the final configuration in~\Cref{fig:mde-nonnested-final-state}, and equivalent sample code for this scenario using the MDE library is provided in~\Cref{code:simpledemo}.

\section{Nested Construction of MDE}
\label{sec:nestedcons}

This section describes a multi-level construction of MDE that we use specifically for points-to sets motivated in~\Cref{sec:introduction}. To further elaborate on the problem, let us consider the sets ${M = \{a \rightarrow \{p, q\}}$, and ${N = \{c \rightarrow \{p, q\}\}}$. In our current construction of MDE, hashing and storing pointer and pointee relations as a single unit is not practically feasible, as their size is not constant. Storing them as points-to pairs is also not reasonable as this does not expose redundancy between pointee sets. Only hashing the pointee sets~\cite{barbar2021hash} does not exploit the redundancy in the entire points-to maps themselves. We therefore need some mechanism to memoize points-to maps at multiple levels, and we do this through the use of \textit{two} \texttt{MDEStates} connected together in a parent-child fashion.

\subsection{Operations on Points-to Maps}
\label{subsec:nestedcons:operdesc}

Before we describe the structure of the augmented MDE, we must first describe how do operations between points-to maps are to be conducted. There are two major categories of operations that are performed in our pointer analysis:
\begin{enumerate}
\item Operations that directly manipulate the graph itself, such as by removing all points-to relations of a given pointer, assigning another pointee set to a pointer and so on.
\item Standard set operations, such as unions, intersections and differences.
\end{enumerate}
In the first category, the points-to map representation has a natural advantage over the equivalent set of points-to pairs representation by being able to directly set or replace the set of pointees for a given pointer, which is not easily expressed in the other representation.

In the second category, we expect the set operations to produce an equivalent result to performing the operation over the set of points-to pairs representation. Let us consider the sets $M$ and $N$ once more. The points-to pair representations of these are ${
	\{a \rightarrow p,
	a \rightarrow q\}
}$ and ${
	c \rightarrow p,
	c \rightarrow q\}
}$ respectively. Thus their union is:
\[\{
	a \rightarrow p,
	a \rightarrow q,
	c \rightarrow p,
	c \rightarrow q
\}\]
The equivalent points-to map representation of the union, given this information is:
\begin{align*}
M \union N &= \{a \rightarrow \{p, q\}\} \union \{c \rightarrow \{p, q\}\} \\
           &= \{
					a \rightarrow \{p, q\},
           			c \rightarrow \{p, q\}
          	  \}
\end{align*}

Observe how in the resultant points-to map, for pointers appearing in only one of the two sets ($a$ and $c$), the pointee sets are same. In such cases, the union operation behaves like a standard set union over atomic elements, which in this case are the pointer-pointee set pairs. 

However, this definition cannot be used if there are pointers appearing in both the sets. Consider two new sets ${J = \{a \rightarrow \{p, q\}, b \rightarrow \{s, t\}\}}$ and ${K = \{b \rightarrow \{t, u\}, c \rightarrow \{p, q\}\}}$. The union between them, in the set of points-to pairs representation is:
\begin{align*}
J \union K &= 
	\{a \rightarrow p,
	  a \rightarrow q,
	  b \rightarrow s,
	  b \rightarrow t 
	\} \union \{
	b \rightarrow t,
	b \rightarrow u,
	c \rightarrow p,
	c \rightarrow q 
\}\\
           &= \{
    a \rightarrow p,
   	a \rightarrow q,
	b \rightarrow s,
	b \rightarrow t,
	b \rightarrow u,
    c \rightarrow p,
    c \rightarrow q
           \}
\end{align*}
The equivalent union under the points-to map representation, given this information, is the following:
\begin{align*}
J \union K &= \{a \rightarrow \{p, q\}, b \rightarrow \{s, t\}\} \union \{b \rightarrow \{t, u\}, c \rightarrow \{p, q\}\} \\
           &= \{
					a \rightarrow \{p, q\},
					b \rightarrow \{s, t, u\},
           			c \rightarrow \{p, q\}
          	  \}
\end{align*}
 To obtain the correct result in the points-to map representation, as inferred from the above two illustrations, we essentially perform a union between the associated pointee sets of $b$ in both $M$ and $N$:
\[
\{s, t\} \union \{t, u\} = \{s, t, u\}
\]
This explicates that the union operation for points-to maps is not ``flat'' in nature, like in the points-to pair set case, but rather operates \textit{recursively}, at two levels of the structure of a points-to map:
\begin{description}
\item[Level 1] performs a standard union over the non-common pointers present in the maps.
\item[Level 2] performs a union over the \textit{pointee sets of individual elements}  of the maps where the pointers are common.
\end{description}
Hence, the union operation is applied to both the structure of the map and its constituent substructures of its elements. Other operations over maps such as intersections and differences can be similarly defined as well. This gives us the main augmented intuition for~\Cref{algo:setunion} we use for a set operation $P \circ Q$ where $P$ and $Q$ are points-to maps:
\begin{enumerate}
\item Let all common pointers between $P$ and $Q$ be $Z$.
\item Let all undivided pointer-pointee set pairs in $P$ with pointers that are not present in $Q$, that is, excluding the pointers in $Z$, to be $P'$.
\item Similarly, let all undivided pointer-pointee set pairs in $P$ excluding pointers in $Z$ be $Q'$.
\item Perform the set operation $P' \circ Q'$ as per its normal semantics. Let the result of this be $R$.
\item Now, iterate over all pointers in Z, obtain its corresponding pointee sets in P and Q, and perform the operation equivalent to `$\circ$' that would give a semantically correct result. Append each result as a pointer-pointee set pair to $R$. This is the final result $R'$.
\end{enumerate}
To illustrate this with our previous example using $J$ and $K$, we first perform a union using the pointers not in common, i.e. $a$ and $c$, with a standard union operation resulting in ${\{a \rightarrow \{p, q\},	c \rightarrow \{p, q\}\}}$. Then we go over the list of common pointers between them, which is only $b$, and we perform the union between the \textit{pointee-sets} of $b$, resulting in the set $\{s, t\}\union\{t, u\}=\{s,t,u\}$. Finally, we append ${b\rightarrow\{s,t,u\}}$ to the previous result, giving us $J\union K=\{a \rightarrow \{p, q\}, b \rightarrow \{s, t, u\}, c \rightarrow \{p, q\}\}$.

\subsection{Augmentations for Nesting}

This model assumes two \texttt{MDEStates}, the parent \texttt{MDEState} storing the entire points-to maps, and a child \texttt{MDEState}, storing only the points-to sets. In the parent, the first major change is that property sets no longer store only properties, but a tuple of the parent property, and an index in the child MDE. We refer to them as the \textit{key} and the \textit{value} respectively. We term this a \textit{property element}. In this way, we once again reduce the elements of a property set to a constant size. The second major change is that we store a reference to the child \texttt{MDEState} state in the parent. We illustrate these changes in~\Cref{fig:nestedmdestate}. There is no augmentation required in the child \texttt{MDEState} itself for enabling the nesting behavior.

\begin{figure}[t]
	\begin{lstlisting}[numbers=none, xleftmargin=0ex]
NestedMDEState<Property, ChildMDE> {
	ChildMDE &ChildState;
	// Property Set Definition:
	using PropertyElement = Tuple<Property, ChildIndex>;
	using PropertySet = Vector<PropertyElement>;
	// Deduplicator objects:
	HashMap<PropertySet*, Index> PropertySetMap;
	Vector<UniquePointer<PropertySet>> Storage;
	// Operations and Maps:
	Union; Intersection; Difference; ... ;
	HashMap<(Index, Index), Bool> SubsetRel;
}
	\end{lstlisting}
	\caption{List of objects defining the state of a nested MDE.}
	\label{fig:nestedmdestate}
\end{figure}

Finally, we augment~\Cref{algo:setunion} with the nesting mechanism in~\Cref{algo:nestedsetunion}. In essence, when we encounter equal keys in two property sets, we simply invoke the same operation (union in this case) on the values of the two property elements in the \textit{child} \texttt{MDEState}, then assign the resultant index to the value field of the result property element. For unequal keys, the respective property elements are operated on trivially by the algorithm, as there are no nested operations required. We may use~\Cref{algo:fulloper} as-is, since the logic remains the same.

\begin{algorithm*}[t]
\caption{Augmented, nested version of~\Cref{algo:setunion}.}
\label{algo:nestedsetunion}
\begin{algorithmic}[1]
\Procedure{NestedUnion}{$ChildState, first, second$}
    \State \textbf{Input:} $ChildState, first, second$
    \State \textbf{Output:} $new\_set$
    \State $new\_set \leftarrow \emptyset$; $cursor_1 \leftarrow first.\text{begin}()$; $cursor\_end_1 \leftarrow first.\text{end}()$
    \State $cursor_2 \leftarrow second.\text{begin}()$; $cursor\_end_2 \leftarrow second.\text{end}()$

    \While{$cursor_1 \neq cursor\_end_1$}
        \If{$cursor_2 = cursor\_end_2$}
            \State \Call{PushRange}{$new\_set, cursor_1, cursor\_end_1$}
            \State \textbf{break}
        \EndIf

        \If{$cursor_2 \rightarrow key < cursor_1 \rightarrow key$}
            \State \Call{PushOne}{$new\_set, *cursor_2$}
            \State $++cursor_2$
        \Else
            \If{$\neg (cursor_1 \rightarrow key < cursor_2 \rightarrow key)$}
                \State $new\_childidx \leftarrow \Call{Operate}{ChildState, Union, cursor_1 \rightarrow value, cursor_2 \rightarrow value}$
                \State $new\_elem \leftarrow (*cursor_1, new\_childidx)$
                \State \Call{PushOne}{$new\_elem, *cursor_1$}
                \State $++cursor_2$
            \Else
                \State \Call{PushOne}{$new\_set, *cursor_1$}
            \EndIf
            \State $++cursor_1$
        \EndIf
    \EndWhile

    \State \Call{PushRange}{$new\_set, cursor_2, cursor\_end_2$}
    \State \Return $new\_set$
\EndProcedure
\end{algorithmic}
\end{algorithm*}

\subsection{Miscellaneous Operations}
Here, we detail some more operations that we utilize in our points-to analysis implementation.

\begin{itemize}
\item \texttt{InsertPointee(Idx, Key, Value)} returns an index to a new set with the pointee-set of \texttt{Key} appended with \texttt{Value}. If \texttt{Key} did not exist in the set previously, it is added in. This is a wrapper around performing a union operation with the current pointee set with a set with \texttt{Value} as the only element.

\item \texttt{UpdatePointees(Idx, Key, ChildIdx)} returns a new index with \texttt{ChildIdx} set as the value for the pointer \texttt{Key} in the property set with index \texttt{Idx}. This does not use an elementary set operation unlike the others mentioned.
\end{itemize}
\subsection{Sample Demonstration}
\label{subsec:nestedcons:operdemo}

Let us consider the previously mentioned pointer relationship sets $J$ and $K$ again. Based on this new arrangement of \texttt{MDEStates}, let us assign the index 1 to $\{p, q\}$, 2 to $\{s, t\}$, and 3 to $\{t, u\}$, and the indices 1 and 2 (in the parent \texttt{MDEState}) to $J$ and $K$ themselves. We thus end up with the following representation. Note how, because of the pointee sets now being represented with indices, it is trivial to determine that $a$ and $c$ point to the same set of pointees:
\begin{align*}
J &= \{a \rightarrow 1, b \rightarrow 2 \} \\
K &= \{b \rightarrow 3, c \rightarrow 1 \}
\end{align*}
Now, in~\Cref{algo:nestedsetunion}, instead of comparing the elements directly, we compare on the \textit{keys}, that is, the pointers, instead of both the pointer and the pointee set. When a key is absent in either set, the algorithm remains the same as in the non-nested case. However, when two keys are the same, we cannot trivially deduce the result of the operation. In this case, we have to invoke the child MDE instance to perform the operation between the pointee sets, to get the required result as described in~\Cref{subsec:nestedcons:operdesc}.

\begin{figure*}[t]
	\centering
	\includegraphics[width=0.8\linewidth]{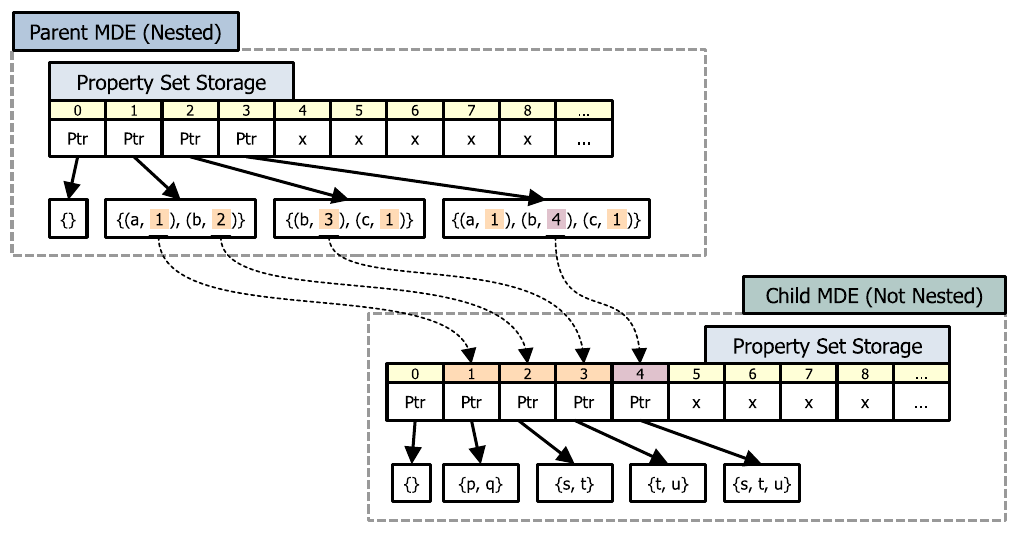}
	\caption{Final configuration of MDEs after peforming the steps in~\Cref{subsec:nestedcons:operdemo}.}
	\label{fig:mde-nested-final-state}
\end{figure*}

Because of $b$, this causes us to perform a union between the pointee-set indices $2$ and $3$. Let us say the new set, $\{s, t, u\}$, maps to the index $4$. Thus the final points-to map $P$ (with say, index 3) becomes:
\begin{align*}
P &= \{a \rightarrow 1, b \rightarrow 4, c \rightarrow 1 \}
\end{align*}
The final state of the nested MDE system is depicted in~\Cref{fig:mde-nested-final-state}, and equivalent sample code for this scenario using the MDE library is provided in~\Cref{code:nesteddemo}.

\subsection{Usage in VASCO-LFCPA and Repurposing of MDE Instances}
\label{subsec:nestedcons:vascousage}
Besides using this nested set of 2 \texttt{MDEStates} for pointer analysis, we would also require the usage of MDE for the representation of liveness analysis data. Instead of trying to create a separate instance of MDE, we actually reuse the child \texttt{MDEState} for pointee-sets in the implementation. This is due to the fact that the structure for data for pointee sets and live variables sets is exactly the same: a set of variables. This allows us to obtain any potential gains in memoization performed in both the pointer analysis and liveness analysis passes of the LFCPA algorithm.

\paragraph*{Extensions and Deeper Insights.}

The nesting mechanism and the repurposing of MDE instances open up doors to potentially useful, interesting and generalized ideas that could be used for applying constructions of MDE to many domains both within and outside compiler design. We detail our final modification of MDE, and the peripheral tools around it in~\Cref{sec:gen-mde}. This is, however, outside of the scope of the points-to analysis part of the paper.

\section{Related Work}
\label{sec:related-work}

Continuing from~\Cref{sec:introduction}, there have been several attempts at improving the data structures used for pointer analysis. This section provides commentary on a selection of works similar, related, or comparable to MDE. We provide an evaluation against MDE for some of these in~\Cref{sec:evaluation}.

\subsection{Earlier Hash-Consing Based Approaches}
A previous work~\cite{barbar2021hash} has made similar observations regarding redundancy in data-flow analysis information, and implemented a similar memoization mechanism to us on the Static Value-Flow Analysis Framework (SVF)~\cite{svf-paper}. Their work was applied on Andersen's analysis~\cite{andersen1994program} and staged flow-sensitive pointer analysis~\cite{hardekopf2011sfs}.

A notable observation to make from this work is that they only de-duplicate the sets of pointees, and not the \textit{entire} points-to representation as a whole. In contrast, our work extends this idea to operate over the entire structure of the data-flow value recursively, allowing sets to be constructed from other de-duplicated values. This enables sharing of common data across multiple levels of representation, rather than only a single component. As a result, our approach can exploit redundancy even when complete equality between undivided data-flow values is absent.

The pointer analysis implementation of the SVF project differs from how VASCO-LFCPA handles data in several significant ways that work to the advantage of this new mechanism, particularly in SVF's use of sparse propagation and per-variable updates. These details are presented in the next section.

\subsection{Internals of the SVF Framework}

The Static Value-Flow Analysis framework~\cite{svf-paper} is a flow- and context-sensitive in nature. It leverages several techniques to make the analysis sparse, that is, reducing the paths along which information needs to be propagated in order to scale. An imprecise and cheap pre-analysis is used to build an approximate interprocedural static-single assignment form of the variables used, and from this, def-use chains of variables are generated. This information is used to select the variables to update at a given program point by an analysis built on top of the framework, such as Andersen's analysis~\cite{andersen1994program}.

The implementation of SVF~\cite{web-svfrepo} also makes use of a significant amount of additional heuristics within its analysis, such as enforcing ``budgets'' on the time consumed by the analysis, the amount of traversals performed in flow- and context-sensitive analysis, which, when exhausted, will make it return conservative approximations for some segments in order to keep the analysis feasbile. This is unlike the current implementation of VASCO-LFCPA, where no such pre-analysis information or heuristics are used. Instead, we depend entirely upon augmentation of the pointer analysis algorithm itself to reduce the amount of information that is processed, and therefore make it cheap.

Also unlike SVF's pointer analysis implementation, data-flow values are operated on in ``bulk'' in VASCO-LFCPA. That is, the entirety of the information at a program point is processed or updated rather than a certain selected variables. It therefore makes sense in SVF to have the pointer relationships stored as a map to a set of pointees, such that the variables to update can be queried directly. This difference makes it very difficult to incorporate MDE into SVF as a replacement for the deduplication infrastructure without an extensive rewrite of the analysis, and it is possible that we may lose the benefits acquired by SVF's method of making the analysis sparse.

\subsubsection{Sparse Bit-Vectors}
\label{subsec:related-work:sbv}
SVF uses LLVM's implementation of a ``sparse'' bit vector (\texttt{llvm::SparseBitVector}~\cite{web-sparsebv, web-llvmhandbook}) as the data structure to store pointee information in SVF. Only the set bits are stored in this structure in sorted order. The bits are stored in a linked list as part of nodes denoting a range within the bit-vector. Variables and other objects thus need to be mapped to a unique integer ID in order to be stored as a bit.

SVF augments the usage of this data structure by creating an internal dense mapping of object IDs to limit the number of linked list nodes that are generated. An internal reverse mapping is created as well to retrieve the actual object IDs later. This mapping operation is expensive, and is done rarely by the analysis.

\paragraph*{Evaluation.} We provide a performance comparison against MDE for this data structure by using it as an alternative for storage of pointee sets in a points-to map in the default implementation. We do not use the dense mapping heuristic used by SVF, however.

\subsection{Binary Decision Diagrams (BDDs)}

Binary Decision Diagrams~\cite[Chapter~6]{huth2004logic} are a way to represent boolean expressions as a tree of true/false decisions (or assignments) on the variables of that expression. BDDs are reduced through several techniques to enable de-duplication of equivalent boolean sub-functions under partial assignments. This allows us to obtain a compact representation of the boolean expression. BDDs are used extensively in formal logic systems to represent and operate upon information.

BDDs have found their use in computing context-sensitive and flow-sensitive pointer analysis information as well~\cite{bddbddb, semisparse-pta-paper} (but not both at the same time) by encoding the entire pointer relationship as a single bit-vector, and the bits subsequently encoded as individual variables of the BDD. Sets of relations are encoded as disjunctions (OR-operations) upon these relationships. Since operations like conjunctions, disjunctions and complements are efficient in BDDs, it makes the pointer analysis cheaper to perform in combination with a few other heuristics.

While this results in an efficient analysis, the generated information is another disjunction of bit-vectors. In other words, the result generated is a logical summary consistent with the set of rules the analysis operates with. Enumerating the corresponding concrete information based on it requires interpreting a large data structure and is not trivial. Our analysis on the other hand, presents concrete information directly.

\subsection{Zero-Suppressed Decision Diagrams (ZDDs)}
\label{subsec:related-work:zdd}
Zero-Suppressed Decision Diagrams~\cite{zddintro} are a variant of BDDs that are particularly useful in representing combinations of sets rather than boolean expressions. ZDDs make their representation compact through similar means as BDDs, but instead store sets of subsets of elements in terms of relationships between those subsets for de-duplication. This too results in a symbolic representation which needs to be traversed to obtain the corresponding concrete information.

Previous work~\cite{zddpta} has used ZDDs for representing pointer relationships. Instead of exposing relationships as large bit-vectors like in the case of BDDs, the work represents them as tuples of a certain number of elements covering a domain (like heap location, type, etc.), and each unique value of an element is represented by a unique ZDD variable. Their approach yielded positive results compared to BDDs for context-sensitive points-to analysis and mixed results in other cases.

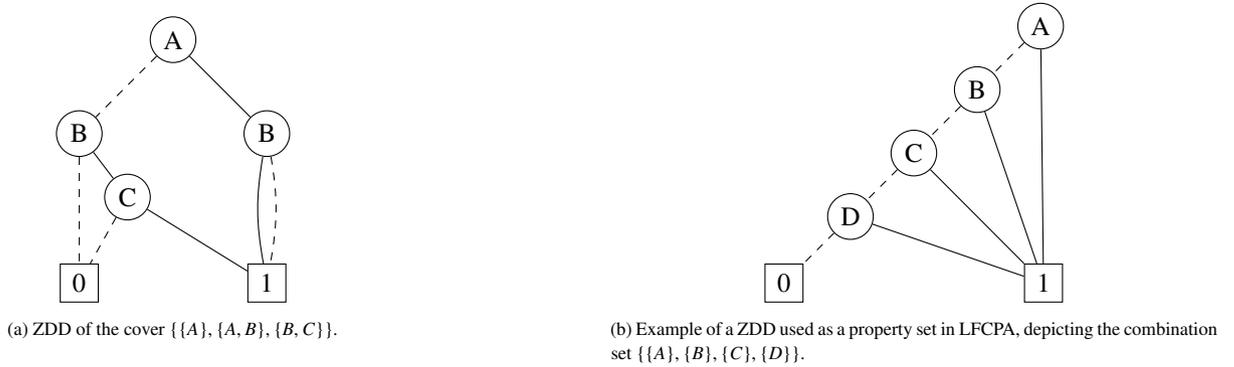
\begin{figure*}[t]
\centering
\begin{subfigure}[t]{0.45\linewidth}
\centering
\begin{tikzpicture}[
    node distance=0.8cm and 0.8cm, 
    roundnode/.style={circle, draw=black, minimum size=6mm, inner sep=1pt},
    terminal/.style={rectangle, draw=black, minimum size=5mm, inner sep=1pt}
]
    \node[roundnode] (a) {A};

    \node[roundnode] (b_low)  [below left=of a] {B};
    \node[roundnode] (b_high) [below right=of a] {B};

    \node[roundnode] (c) [below right=0.4cm and 0.2cm of b_low] {C};

    \node[terminal] (zero) [below=1.4cm of b_low] {0};
    \node[terminal] (one)  [below=1.4cm of b_high] {1};

    \draw (a) -- (b_high);
    \draw[dashed] (a) -- (b_low);

    \draw (b_low) -- (c);
    \draw[dashed] (b_low) -- (zero);

    \draw (b_high) to[bend right=10] (one);
    \draw[dashed] (b_high) to[bend left=10] (one);

    \draw (c) -- (one);
    \draw[dashed] (c) -- (zero);

\end{tikzpicture}
\caption{ZDD of the cover $\{\{A\}, \{A, B\}, \{B, C\}\}$.}
\label{subfig:zddconst:normal}
\end{subfigure}
\hfill
\begin{subfigure}[t]{0.45\linewidth}
\centering
\begin{tikzpicture}[
    node distance=0.8cm and 0.8cm,
    roundnode/.style={circle, draw=black, minimum size=6mm, inner sep=1pt},
    terminal/.style={rectangle, draw=black, minimum size=5mm, inner sep=1pt}
]

  \node[roundnode] (A) {A};
  \node[roundnode] (B) [below left=0.4cm and 0.4cm of A] {B};
  \node[roundnode] (C) [below left=0.4cm and 0.4cm of B] {C};
  \node[roundnode] (D) [below left=0.4cm and 0.4cm of C] {D};

  \node[terminal] (zero) [below left=0.4cm and 0.4cm of D, terminal] {0};

  \node[terminal] (one) [right=2.9cm of zero, terminal] {1};

  \draw [dashed] (A) -- (B);
  \draw [dashed] (B) -- (C);
  \draw [dashed] (C) -- (D);
  \draw [dashed] (D) -- (zero);

  \draw (A) -- (one);
  \draw (B) -- (one);
  \draw (C) -- (one);
  \draw (D) -- (one);

\end{tikzpicture}
\caption{Example of a ZDD used as a property set in LFCPA, depicting the combination set $\{\{A\}, \{B\}, \{C\}, \{D\}\}$. } 
\label{subfig:zddconst:lfcpa}
\end{subfigure}

\caption{Examples of ZDDs. Nodes are ordered on a lexicographical basis. Traversal along a solid edge or dotted edge implies presence or absence of an element respectively. Note how, in \Cref{subfig:zddconst:lfcpa} the left spine resembles a sorted vector or linked list.}
\label{fig:zddconst}
\end{figure*}

\paragraph*{Evaluation.} We provide a performance comparison against MDE for this data structure. Unlike in the aforementioned paper, we only use them as a replacement for the pointee sets in a points-to map in the default implementation by representing it as a set of disjoint single-element subsets. This results in a structure that is very similar in appearance and function to a linked list or a sorted vector like the one we use in MDE. We provide a visual representation of this configuration in~\Cref{fig:zddconst}.

We implement this replacement through the use of the Colorado University Decision Diagram (CUDD) package~\cite{cudd-package}. It is worth noting that CUDD implements de-duplication and operation memoization in a very similar way to a non-nested or single-level MDE, and thus benefits the analysis in similar ways as well.

\subsection{Summary}
From the above approaches, it can be seen that optimizations on the representation of the data in an analysis have been substantially attempted before. However, the existing techniques either operate at the level of whole sets such as with BDDs and ZDDs, or only focus on the particulars such as pointee-sets in the case of the hash-consing based approach.

Our approach in contrast specifically targets the structural redundancy at multiple levels of the pointer relationships. This new representation also potentially plays into the advantage of analyses like VASCO-LFCPA, which make a liberal use of data-flow values within the analysis algorithm.

\section{Evaluation}
\label{sec:evaluation}

Here, we provide an evaluation of MDE against the following configurations of the data representation in VASCO-LFCPA:

\begin{enumerate}
\item The default implementation with \texttt{std::set} for pointee sets and \texttt{std::map} for the points-to map.
\item A \textit{single-level} implementation with MDE used to represent the pointee-sets and \texttt{std::map} for the points-to map.
\item A multi-level implementation with MDE as detailed in~\Cref{subsec:nestedcons:vascousage}.
\item An implementation with sparse bit-vectors as detailed in~\Cref{subsec:related-work:sbv}.
\item An implementation with ZDDs as detailed in~\Cref{subsec:related-work:zdd}.
\end{enumerate}

We measure the time taken and the peak memory usage in each case for the SPEC~\cite{speccpu2006} Benchmarks \texttt{lbm}, \texttt{mcf}, \texttt{libquantum}, \texttt{bzip2}, \texttt{sjeng}, and \texttt{hmmer}. Evaluation materials and source code for all three of these claims are provided as part of an artifact~\cite{mdeartifact}, except for the SPEC Benchmarks~\cite{speccpu2006}, due to them being licensed software.

\paragraph*{Experimental Setup.} We use a virtual machine configured with Ubuntu 22.04, 4vCPUs, 64 GB of RAM, and 75 GB of disk space. This VM is provisioned on a university-maintained data center. The physical node it is hosted on has two 10-core CPUs and 256 GB of RAM. We use the Linux \texttt{time} application to measure lifetime peak memory usage, and an in-code timer to measure analysis time. As a part of the artifact, we provide a Docker image and a Dockerfile with Ubuntu 24.04 for the system configuration as a part of our artifact. All programs are compiled with either GCC 13.3.0 (in Ubuntu 24.04) or 11.4.0 (in Ubuntu 22.04). The optimization flag \texttt{-Ofast} is passed to the compiler for the build of the VASCO-LFCPA implementation. For all configurations, we take the arithmetic mean of three observations except for the \texttt{hmmer} and \texttt{sjeng} because of high time consumption. For these, we only provide the mean of two observations instead.

\paragraph*{The VASCO-LFCPA Implementation.}
We have integrated MDE on a C++ implementation of VASCO-LFCPA, which is a work-in-progress effort. It is implemented on top of LLVM~14's infrastructure, with an in-house abstraction layer called SLIM (Simplified LLVM IR Modeling). VASCO itself is implemented as a separate, generic module and is used as a foundation to implement LFCPA. The implementation has a large number of considerations made for both the theoretical aspects of the analysis and the practical aspects of LLVM's program representation. Reimplementing this analysis from scratch is infeasible in a reasonable time frame and is beyond the scope of this work. Instead, we present this evaluation as an observation of the performance of MDE within a realistic and representative setting, thus demonstrating its impact under practical constraints. We provide more details regarding the complexity of the VASCO-LFCPA implementation in~\Cref{sec:vasco-complexity}.

Due to the usage of non-deterministic data structures (like \texttt{std::unordered\_map}) and the program objects being internally represented with pointers, statistics such as instructions processed and operations performed are not constant with multiple runs on the same input. They are, however, consistent within a certain range of error, hence we provide the mean of multiple observations.

The application currently only supports LLVM IR files created from C programs. For many real-world programs the implementation also requires the corresponding LLVM IR files used as input to be given a set of special transformations in order to accomodate features not covered by it, such as decomposition of multi-argument \texttt{GetElementPtr}~\cite{web-gepinst} instructions, constant expressions in instructions, and so on. The SPEC benchmark programs are the only ones where there has been substantial testing with this implementation, and hence are the ones used for evaluation.

To accommodate the sparse bit-vector and ZDD-based implementations, the code has been modified to add a mapping routine that assigns a unique integer identifier to each variable object present in VASCO-LFCPA, which are represented using pointers. Adding a mapping is a one-time cost and subsequent look-ups to the same pointer or integer have no additional cost besides the look-up itself.

\subsection{Results}

The results of running the analysis in each case are detailed in~\Cref{tab:resultspmu} and~\Cref{tab:resultstime}. For the smaller benchmarks, namely \texttt{lbm} and \texttt{mcf}, the single-level approach comes up as the most efficient one. We explain the poor performance of the multi-level MDE implementation in the case of \texttt{mcf} by the high amount of total unique sets, and the disproportionately low number of equality comparisons as shown in~\Cref{tab:lfcpa-comparison}, suggesting that the overhead of registering the sets outweighs the efficiency gains from the memoization. In the rest of the results, however, the multi-level MDE-based implementation comes up as the most efficient one by large margins. In the benchmark \texttt{hmmer}, the reduction in runtime is as high as \timeimprovement{} compared to the default implementation.

In terms of peak memory consumption, multi-level MDE comes up as the most efficient in all of the cases with the increase in the size of input (as per~\Cref{tab:lfcpa-comparison}). Compared to the default implementation, the multi-level MDE implementation consumes a fraction of memory at peak. In \texttt{hmmer}, the reduction is as high as \memimprovement{} compared to the default implementation.

Hence, there is empirical evidence pointing to MDE being useful in applications like these. In general, from the results of the ZDD-based implementation and the single-level MDE implementations, while less performant than the multi-level MDE implementation, point to memoization being a strongly contributing component for efficiency in the analysis. We present further inferences from this data in~\Cref{sec:insights}.

\begin{table*}[t]
\begin{subtable}{\linewidth}
	\centering
    \begin{tabular}{lrrrrr}
        \toprule
        \textbf{Benchmark} & \textbf{Default} & \textbf{Sparse Bit-Vector} & \textbf{ZDD} & \textbf{Single-level MDE} & \textbf{Multi-level MDE} \\
        \midrule
\texttt{lbm}
	& \SI{77}{\mega\byte} 
	& \SI{75}{\mega\byte}  (1.02x)
	& \SI{166}{\mega\byte} (0.46x)
	& \SI{70}{\mega\byte}  (1.10x)
	& \SI{67}{\mega\byte}  (1.15x)\best 
	\\

\texttt{mcf}
	& \SI{481}{\mega\byte} 
	& \SI{322}{\mega\byte} (1.50x)
	& \SI{258}{\mega\byte} (1.86x)
	& \SI{193}{\mega\byte} (2.49x)
	& \SI{102}{\mega\byte} (4.71x)\best 
	\\

\texttt{libquantum}
	& \SI{163}{\mega\byte} 
	& \SI{145}{\mega\byte} (1.12x)
	& \SI{211}{\mega\byte} (1.41x)
	& \SI{116}{\mega\byte} (1.41x)
	& \SI{100}{\mega\byte} (1.63x)\best 
	\\

\texttt{bzip2}
	& \SI{12.28}{\giga\byte} 
	& \SI{7.27}{\giga\byte} (1.69x)
	& \SI{3.78}{\giga\byte} (3.24x)
	& \SI{3.53}{\giga\byte} (3.48x)
	& \SI{849.50}{\mega\byte} (14.34x)\best 
	\\

\texttt{sjeng}\approxdata
	& \SI{47.11}{\giga\byte} 
	& \SI{37.88}{\giga\byte} (1.24x)
	& \SI{15.77}{\giga\byte} (2.99x)
	& \SI{15.41}{\giga\byte} (3.06x)
	& \SI{2.61}{\giga\byte}  (\memimprovement{})\best 
	\\

\texttt{hmmer}\approxdata
	& \SI{27.41}{\giga\byte} 
	& \SI{16.62}{\giga\byte}(1.65x)
	& \SI{7.95}{\giga\byte} (3.45x)
	& \SI{7.84}{\giga\byte} (3.50x)
	& \SI{2.74}{\giga\byte} (10.01x)\best 
	\\
        \bottomrule
    \end{tabular}
    \caption{Experimental Results of peak memory usage (PMU) of analysis in different configurations.}
    \label{tab:resultspmu}
\end{subtable}

    \vspace{1em}

\begin{subtable}{\linewidth}
    \centering
    \sisetup{scientific-notation = true}
    \begin{tabular}{lrrrrr}
        \toprule
        \textbf{Benchmark} & \textbf{Default} & \textbf{Sparse Bit-Vector} & \textbf{ZDD} & \textbf{Single-level MDE} & \textbf{Multi-level MDE}  \\
        \midrule
\texttt{lbm}
	& \SI{227}{\milli\second}        
	& \SI{217}{\milli\second} (1.05x) 
	& \SI{861}{\milli\second} (0.26x) 
	& \SI{192}{\milli\second} (1.18x)\best 
	& \SI{196}{\milli\second} (1.16x)
	\\

\texttt{mcf}
	& \SI{27251}{\milli\second} 
	& \SI{23037}{\milli\second}(1.18x) 
	& \SI{19880}{\milli\second} (1.37x) 
	& \SI{18004}{\milli\second} (1.51x)\best 
	& \SI{30301}{\milli\second} (0.90x)
	\\

\texttt{libquantum}
	& \SI{5.9e3}{\milli\second} 
	& \SI{5.1e3}{\milli\second} (1.16x)
	& \SI{4.8e3}{\milli\second} (1.23x)
	& \SI{3.9e3}{\milli\second} (1.51x)
	& \SI{3.4e3}{\milli\second} (1.74x)\best 
	\\

\texttt{bzip2}
	& \SI{2.994e6}{\milli\second} 
	& \SI{2265499}{\milli\second} (1.36x)
	& \SI{1787909}{\milli\second} (1.72x)
	& \SI{1582692}{\milli\second} (1.94x)
	& \SI{1385876}{\milli\second} (2.22x)\best 
	\\

\texttt{sjeng}\approxdata
	& \SI{28799908}{\milli\second} 
	& \SI{26217504}{\milli\second} (1.10x)
	& \SI{18170776}{\milli\second} (1.58x)
	& \SI{16977782}{\milli\second} (1.70x)
	& \SI{12201986}{\milli\second} (2.36x)\best 
	\\

\texttt{hmmer}\approxdata
	& \SI{45504510}{\milli\second} 
	& \SI{29528985}{\milli\second} (1.54x)
	& \SI{17259587}{\milli\second} (2.64x)
	& \SI{14923989}{\milli\second} (3.05x)
	& \SI{5581260}{\milli\second}  (\timeimprovement{})\best 
	\\
        \bottomrule
    \end{tabular}
    \sisetup{scientific-notation = false}
    \caption{Experimental Results of time consumtion of analysis in different configurations.}
    \label{tab:resultstime}
\end{subtable}
\begin{tablenotes}
\item \best Best statistic for the given benchmark.
\item \approxdata Data is collected from only two runs due to high time consumption.
\end{tablenotes}
    \caption{Experimental results of running the VASCO-LFCPA analysis implementation on selected SPEC benchmark programs. Performance gain factor with respect to the default implementation is shown to the right of each entry in parantheses. Each data point is mean of three runs.}
\label{tab:results}
\end{table*}

\subsection{Additional Metrics and Observations}
\label{subsec:gen-mde:additional-metrics}
The run-time behavior of MDE can be further examined to obtain several interesting statistics. These would allow us to judge the quality of the memoization being done for each operation in an analysis program for a given input of configuration. We devise the following metrics to judge various aspects of the memoization, which closely resemble metrics that are used to evaluate caching mechanisms: 
\begin{description}
\item[Hits:] Operation is already memoized.
\item[Equal Hits:] Operands are equal ($a \equiv b$).
\item[Subset Hits:] One operand is a subset of another operand.
\item[Empty Hits:] One of the operands is an empty set.
\item[Cold Misses:] The result of the operation does not exist: $c$ in $a \circ b = c$.
\item[Edge Misses:] The operation does not exist in the map, but the result exists: $a \circ b$ in $a \circ b = c$.
\end{description}
The rationale behind the usage of the term ``edge miss'' is provided in~\Cref{subsec:gen-mde:parallels}, which alludes to parallels between MDE and abstract mathematical structures.

Sample statistics for these metrics against various benchmarks for the union operation are shown in~\Cref{tab:sample-statistics}. They are generated under the same configuration described in the previous experimental setup, but all analyses are ran seven times instead of three. As can be seen, the hit ratios are greater than $90\%$ for all of the benchmarks, implying a high amount of redundancy within the analysis. The comparatively high number of empty hits and equal hits within the benchmarks heavily imply that data within the analysis is in fact sparse and duplicated. Programs using MDE should aim to improve the hit ratio in order to increase their performance, as alluded to by the comparatively poorer ($<95\%$) hit ratios in the \texttt{lbm} and \texttt{mcf} benchmarks.

However these metrics, hypothetically, can also be representative of the amount of redundant data in the input payload, with the number of hits being proportional to redundancy. The presence of a high number of cold and edge misses could imply that there is too much entropy in the input data and that redundancy is scarce. This line of thought remains underexplored. There is potentially substantial scope for such information-theoretic experimentation with MDE, and may be a direction for future work with this mechanism.

\begin{table}[t]
\centering
\begin{tabular}{llrrrrrr}
\toprule
 & & \textbf{\texttt{lbm}} & \textbf{\texttt{mcf}} & \textbf{\texttt{libquantum}} & \textbf{\texttt{bzip2}} & \textbf{\texttt{sjeng}} & \textbf{\texttt{hmmer}} \\
\midrule
\multirow{6}{*}{\textbf{Pointee Sets}}
 & \textbf{Hits}        & \num{2035    } & \num{101047  } & \num{55156   } & \num{2187426 }  & \num{7509425 } & \num{16445316}  \\
 & \textbf{Equal Hits}  & \num{3026    } & \num{298816  } & \num{14773   } & \num{2891014 }  & \num{2708378 } & \num{4156643 }  \\
 & \textbf{Subset Hits} & \num{1860    } & \num{41548   } & \num{15763   } & \num{1160364 }  & \num{400054  } & \num{1675858 }  \\
 & \textbf{Empty Hits}  & \num{3660    } & \num{69005   } & \num{66878   } & \num{1116770 }  & \num{5722175 } & \num{4060769 }  \\
 & \textbf{Cold Misses} & \num{931     } & \num{12556   } & \num{3278    } & \num{73643   }  & \num{22311   } & \num{431247  }  \\
 & \textbf{Edge Misses} & \num{38      } & \num{7727    } & \num{904     } & \num{20114   }  & \num{6695    } & \num{138137  }  \\
 & \textbf{Hit Ratio}   & 91.61\%        & 96.18\%        & 97.33\%        & 98.74\%         & 99.82\%        & 97.88\%         \\
\midrule
\multirow{6}{*}{\textbf{Points-to Maps}}
 & \textbf{Hits}        & \num{414     } & \num{57933   } & \num{35259   } & \num{1486177 }  & \num{5457418 } & \num{5809949 }  \\
 & \textbf{Equal Hits}  & \num{9729    } & \num{75797   } & \num{112479  } & \num{1427804 }  & \num{9125807 } & \num{4772366 }  \\
 & \textbf{Subset Hits} & \num{1914    } & \num{44572   } & \num{36137   } & \num{1172046 }  & \num{5599226 } & \num{3154907 }  \\
 & \textbf{Empty Hits}  & \num{5960    } & \num{33550   } & \num{69973   } & \num{840800  }  & \num{4769579 } & \num{3083344 }  \\
 & \textbf{Cold Misses} & \num{5       } & \num{1395    } & \num{40      } & \num{24347   }  & \num{9540    } & \num{91680   }  \\
 & \textbf{Edge Misses} & \num{1073    } & \num{21071   } & \num{1696    } & \num{234591  }  & \num{201474  } & \num{450966  }  \\
 & \textbf{Hit Ratio}   & 94.35\%        & 90.41\%        & 99.32\%        & 95.01\%         & 99.16\%        & 96.87\%         \\
\bottomrule
\end{tabular}
\vspace{1em}
\caption{Sample statistics for the union operation for selected SPEC benchmarks from the MDE-based configuration for VASCO-LFCPA (see~\Cref{sec:evaluation}). Data is mean of 7 observations.}
\label{tab:sample-statistics}
\end{table}

\section{Insights}
\label{sec:insights}

\paragraph*{Eventual Saturation.} We would like to emphasize the trend that is shown for MDE by the data presented in \Cref{tab:results}. With the increase in the size of the input program, both the analysis time and the memory performance factors increase with respect to the default implementation. While the size itself does not solely determine analysis cost, this may suggest that information generated in the analysis starts to saturate after some threshold within each program.

\paragraph*{Overheads and Multi-level Deduplication.} The results also suggest that performing de-duplication recursively is an effective method of introducing efficiency into the analysis despite the overheads incurred by the process. Exploration into devising an encoding of data in a manner that exposes the largest amount of redundancy in it is therefore a viable strategy.

\paragraph*{Closed-World Computation.} We would also like to point out the fact that for any given input program, memory consumed by the MDE-based implementation remains consistently less than the default one. This is despite the fact that none of the information stored in MDE is ever discarded for the duration of the analysis unlike in the default case where the data representation is mutable. While we have already pointed out redundancy in data and operations as a reason for this, it may not necessarily mean that given an infinite amount of time, MDE will not consume an infinite amount of memory as the system receives data. Our observations run contrary to this intuition.

A more fundamental reasoning for this is the fact that the only input given to the analysis is the program itself. All information synthesized in the analysis comes solely from the input program. We term such processing of information as \textit{closed-world computations}.

From this, it follows that the amount of unique information that can be synthesized from the input, given that no external information is introduced during the process of synthesis, is finite. Therefore, redundancy in the synthesized information will eventually be encountered given enough time as the analysis progresses, creating opportunity for reuse of data even if the original data itself has a low amount of redundancy.

This observation helps explain why MDE does not show unbounded growth in memory during runtime despite data being immutable in nature, and why being able to memoize data, especially in long-running analyses like whole-program pointer analysis has been effective in increasing performance in our observations.

However, the above intuition holds only when the analysis itself is guaranteed to terminate. A counter example to this is value range analysis~\cite{rnganalysis-paper}, where we attempt to statically determine the quantitative range of values a variable can take. Unless widening techniques~\cite[chapter 33]{cousot2021principles} are applied to such an analysis to guarantee its termination, MDE may be of little use in a situation like this. It will eventually cause unbounded growth in its memory consumption due to the analysis perpetually generating new ranges, which can extend up to negative and positive infinity. In other words, MDE cannot help in ensuring termination when an analysis computes unbounded data.

\paragraph*{Relation to VASCO-LFCPA.} Continuing from~\Cref{sec:related-work}, we believe the analysis presented by VASCO-LFCPA, compared to others such as the one presented by SVF, stands to benefit more from the multi-level mechanism given by MDE. While SVF aims to make the analysis efficient through the use of preanalysis and narrowing down propagation paths, VASCO and LFCPA, on the other hand, use the generated data-flow values themselves in an effort to make the analysis more efficient. VASCO, in particular, has to use data-flow values to distinguish interprocedural contexts. Without a system like MDE, comparisons on data-flow values have to be expensive non-constant time operations, causing a significant bottleneck. As illustrated in~\Cref{tab:lfcpa-comparison}, a large percentage of operations in an analysis are just equality comparisons.

\section{Future Work}
\label{sec:future-work}
While the application of MDE shows a substantial benefit to our existing pointer analysis efforts, the current integration with the pointer analysis implementation may not necessarily be optimal due to the inherent complexity of the analysis code and needs to be revised. There is still a lot more room and potential for further optimization, both in the algorithmic and data structure fronts. The runtime of the analysis, while cut down significantly, needs to be further brought down to more reasonable durations. MDE is currently implemented on top of standard C++ STL data structure implementations. Switching to more performant alternatives such as the ones from the Boost~\cite{web-boost} or Abseil~\cite{web-abseil} C++ libraries, or using custom allocators and allocation patterns may yield more efficiency in MDE.

Other possible areas of optimization may include identifying more areas of capturing redundancy using more alternate constructions of MDE, or modifying the algorithm to utilize the redundancy more effectively. In the future, we aim to find more insights into representational optimizations like MDE for creating performant analyses. We detail the final, generalized structure of MDE we aim to use for such exploration in~\Cref{sec:gen-mde}. Moreover, the information-theoretic aspects of MDE need to be explored further in order to obtain a clear correlation between an input program and the redundancy within it, potentially helping us discover further approaches of optimization.

\section{Conclusion}
\label{sec:conclusion}

In this paper we have presented Multi-level Deduplication Engine (MDE), a mechanism that performs recursive de-duplication and assigns unique identifiers to data items as the basis for building the data structure used for the representation of data-flow values. Our approach allows de-duplicated data items to themselves be constructed from other de-duplicated items, enabling a recursive structure in which sharing occurs across multiple levels of representation. This multi-level structure enables reuse of data-flow value instances, and reuse even when equality between data-flow values as a whole is absent, capturing structural redundancy that single-level approaches cannot exploit. MDE helps in achieving efficiency without the need of changing the analysis algorithm substantially.

We provide a full C++ implementation of MDE, and have compared MDE's integration of our analysis against several alternatives based on previous work. Evaluation on standard benchmarks shows a reduction up to \memimprovement{} in peak memory usage and \timeimprovement{} in runtime. The application exhibits an upward trend of effectiveness with the increase in benchmark size.

Our main goal with MDE is to create a solution that allows us to perform whole-program analysis efficiently. MDE aims to make it so that the availability of primary memory is no longer a constraining factor of the analysis. While MDE has already shown us substantial gains towards achieving these goals, there is much potential for further improvements, both in terms of providing a more optimal integration within the analysis, and through sophisticated constructions of data representations and transactions in analyses. In the future, we hope to find more applications of MDE beyond compiler design and optimization.

\bmsection*{Financial disclosure}

None reported.

\bmsection*{Conflict of interest}

The authors declare no potential conflict of interests.
\bibliography{BIBLIOGRAPHY}

\bmsection*{Supporting information}

The data that supports the findings of this study are openly available in Zenodo at \url{https://doi.org/10.5281/zenodo.19437315}, reference number 19437315.

\appendix

\section{Generalizing MDE}
\label{sec:gen-mde}

This, and the following sections provide a description of a general modeling of MDE that naturally emerges from the previous description provided for pointer analysis. This is done in the hopes that it would be useful for other program analysis problems, and possibly, for problems outside the domain of program analysis. This modeling allows for an arbitrarily nested structure in the form of multiple child MDEs, and thus be used for modeling arbitrary data representations that contain aggregate data in the form of sets, maps and so on.

\subsection{Parallels with Mathematical Lattices}
\label{subsec:gen-mde:parallels}

One of the interesting parallels that emerges from the way data in MDE is structured is that it in some ways resembles a lattice that is incrementally being built from the ground up in certain cases.

Consider the union operation once more. The data that is stored for each memoized operation performed essentially describes a lattice join operation between two sets. As more operations are performed, we incrementally record the join operations between the sets. It can be described as the bottom-up formation of non-unique deriviation trees from a set to its subsets. These trees, when merged together, essentially form a part of the join-semilattice of analysis data under the union operation. Similar commentary can be made for the intersection operation. The subset map found in MDE forms an additional way to record joins in the join-semilattice, and the same data can be repurposed as meets in the meet-semilattice under the intersection operation.

\begin{figure*}[t]
	\centering
	\includegraphics[width=0.7\linewidth]{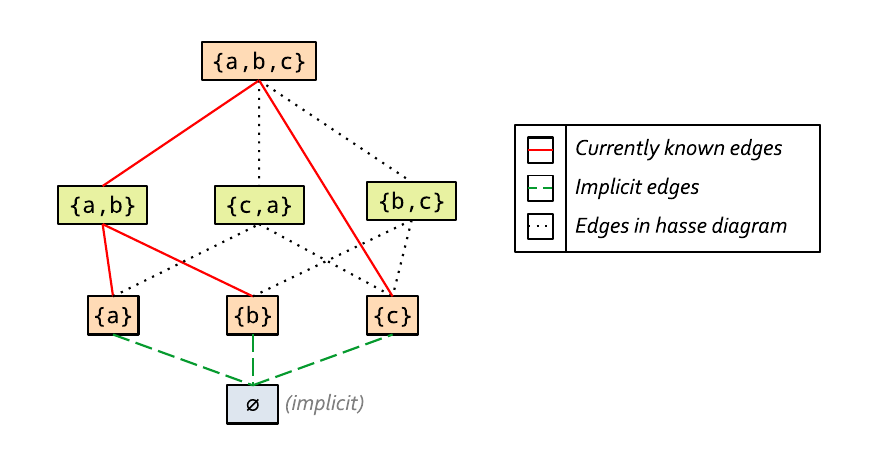}
	\caption{Gradual, incremental lattice structure emerging from the usage of MDE from the union operation. Note how the red edges form a tree, but it is not the only possible tree leading to the top-most node ($\{a, b, c\}$).}
	\label{fig:gen-mde-lattice}
\end{figure*}

If the information discovered in an analysis is imagined as a hasse diagram~\cite{wiki-hassediag}, we are essentially recording all of the edges of relationships in a hasse diagram, including the redundant ones that are not usually shown. An illustration of this is provided in~\Cref{fig:gen-mde-lattice}.

Drawing this parallel, we arrive at the reason behind the usage of the term ``edge miss'' in~\Cref{subsec:gen-mde:additional-metrics}. We use it to denote the lack of edges in the hasse diagram representation from the operands to the required result. Both the operands and the result are present within the hasse diagram, but the relationship between them remains unknown until the corresponding operation is actually performed.

\section{General Construction and Usage of MDE}
\label{sec:general}
\begin{figure*}[t]
	\centering
	\includegraphics[width=0.8\linewidth]{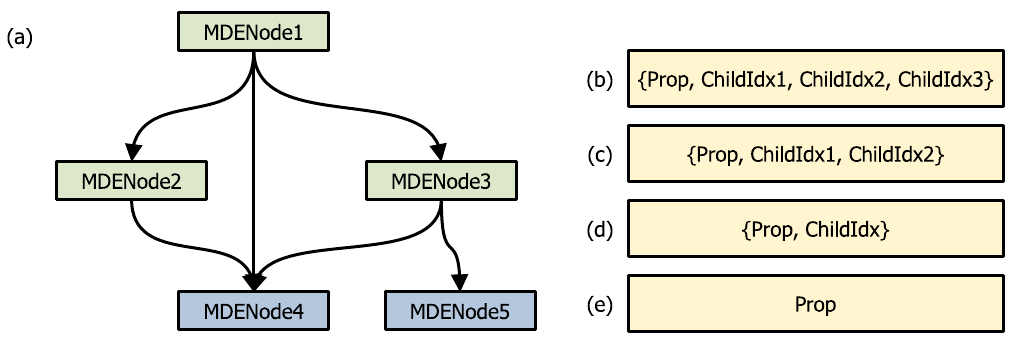}
	\caption{A network of generalized MDE nodes. (b), (c), (d) depict the property elements of node 1, node 3, and node 2 respectively. (e) depicts the property elements of the non-nested nodes 4 and 5, which are the ``leaves'' in the network.}
	\label{fig:gen-mde-network}
\end{figure*}

We deviate in two important ways from the previous formulation of MDE to provide the generalized structure. The first is that we extend the notion of a parent and child \texttt{MDEStates} to form a directed acyclic graph (DAG)-like construction of MDE. Henceforth, we term each \texttt{MDEState} to be a \textit{node} in the construction of the complete MDE. Each MDE node can interact recursively with its children in order to store and operate on data. By extension, the property elements used by the MDE nodes now can include an arbitrary number of nested indices based on the number of child nodes the current MDE node has. An example of this set-up is illustrated in~\Cref{fig:gen-mde-network}. Like in the previous formulation, MDENodes can have custom-defined implementations on how each operation should be performed. We consider the name of the operation to be its sole canonical identifier in all nodes. Hence, recursive calls for an operation, if required, are performed by calling the function that has the name of the operation on each respective child node. To illustrate the generalized nature of the MDE implementation we depict a two-child configuration of MDE in~\Cref{code:2elemdemo}.

Note how this changes the type signature of each participating node. A node may store sets of integers, floats, or other types as its elements, but at the same time it may also store nested references to nodes storing sets of different data types.

\paragraph*{Implementation Details.} The construction of this network of nodes while managing the variation of types of each node is achieved statically through the heavy use of template and preprocessor metaprogramming in the C++ implementation. Using template parameters, we decide on whether a MDE node is nested or not, and if it is nested, what child MDE nodes are to be supplied as references during initialization. Similarly for the standard operations (union, intersection, etc.) we also describe statically how the nested operation is to be performed using compile-time (\texttt{constexpr}) template switches as per the method described earlier in~\Cref{sec:nestedcons}, but generalized to an arbitrary number of child elements.

\subsection{Context-Free Grammar of MDE Constructions}
\label{sec:gen-mde-cfg}
This generalized nested construction of an MDE can be expressed with notation that forms a context-free grammar. An illustration of the grammar is provided in~\Cref{fig:lhf-cfgrammar}. In this notation, we define a node in terms of parameters: \textit{Prop}, the domain of sets of a given property (essentially the power set of the property's domain), \textit{OP}, the set of operations that can be performed on that domain, finally followed by a list of optional ordered parameters called \textit{Nesting}, describing the child MDE nodes recursively. For example, MDEs for the points-to map and the set of live variables, respectively, can be described as:
\begin{figure*}[t]
\begin{minipage}{\linewidth}
\[
\begin{aligned}
\nonumber\textbf{MDEConstr} &\prodarrow \text{MDENode}(\textbf{Prop}, \textbf{OP}, \textbf{Nesting}) \rsep \text{MDENode}(\textbf{Prop}, \textbf{OP})\\
\nonumber\textbf{Nesting} &\prodarrow \textbf{MDEConstr}, \textbf{Nesting} \rsep \textbf{ MDEConstr} \\
\nonumber\textbf{Prop} &\prodarrow \text{<identifier>} \rsep \text{<set of property sets>} \\
\nonumber\textbf{OP} &\prodarrow \text{<identifier>} \rsep \text{<set of operations>}
\end{aligned}
\]
\vspace{1pt}
\end{minipage}
\caption{Illustration of a context-free grammar for MDE constructions.}
\label{fig:lhf-cfgrammar}
\end{figure*}
\[
\begin{aligned}
\nonumber\textit{PTGraphMDE} &=
	\text{MDENode}(\textit{var}, \textit{normalop}, \text{MDENode}(\textit{var}, \textit{normalop})) \\
\nonumber\textit{LiveMDE} &= \text{MDENode}(\textit{var}, \textit{normalop})
\end{aligned}
\]
Where \textit{var} is all of the possible sets of variable objects, and \textit{normalop} is a set of standard operations consisting of union, intersection, difference, etc. The property element of the respective MDE is derived from the net effect of the notation. Such as in the case of \textit{PTGraphMDE}, we know that the property elements will contain an index to a child set alongside the property itself due to the nesting parameter.

The significance of having this notation comes from optimizations on a construction of MDE we may be able to perform with it. Note how the nesting parameter of \textit{PTGraphMDE} is exactly the same as the description of \textit{LiveMDE}. Hence, the resultant MDENode can be reused in \textit{PTGraphMDE}, reducing \textit{PTGraphMDE} to the following:
\[
\nonumber\textit{PTGraphMDE} =
	\text{MDENode}(\textit{var}, \textit{normalop}, \textit{LiveMDE})
\]
We made this observation informally earlier in \Cref{subsec:nestedcons:vascousage}, however, this context-free grammar-based notation gives us a methodical way to deduce such optimizations, as well as turn it into an automated process. We have implemented an experimental Python program provided as part of the artifactnote{Present in \texttt{src/builder\_tool/} in the MDE source code directory.} that can convert a JSON-based declaration into a set of C++ declarations, however it does not capture all of the possible edge-cases that can occur within the notation.

Another configuration the notation allows us to express is that of an MDE node containing sets of indices present in another MDE node, as a form of indirect or implicit nesting. Essentially, this forms sets of integers as indices are integers themselves, and does not require any special consideration for MDE. Such a configuration may potentially turn out to be useful in certain cases. We describe a possible use case in~\Cref{subsec:additional-motiv:gpg}.


\subsection{Extending the MDE Implementation}

The entire state of an MDE node and operations on property sets are contained within a single class in the C++ implementation. This class can be extended to support the users needs through the implementation of additional operations and helper functions. However, common routines such as that of deducing trivial operations or new registrations in the case of a miss are not consolidated as common, reusable blocks currently at the time of writing, making the implementations slightly difficult. We hope to add a more streamlined flow for adding operations in the future.

\section{Additional Features in the C++ Implementation of MDE}

Besides the features described in this paper, the C++ implementation includes several other in progress, experimental features that can be potentially useful in performing efficient analysis.

\paragraph*{Parallelization.} The current implementation of MDE can be enabled to have thread safety at compile-time. That is, data and operations can be performed without affecting the integrity of the mechanism. There are two implementations available, one using the POSIX threading interface~\cite{wiki-pthreads}, and the other using the Intel Thread Building Blocks (TBB) parallelization library~\cite{web-inteltbb}. The availability of multithreading opens the door to features like optimistic pre-processing using worker threads to increase analysis efficiency, or in cases where the entire analysis can be run in parallel.

\paragraph*{Eviction.} Even though we consider MDE to perform well in terms of memory efficiency, there may be situations where the user may know beforehand if a certain collection of property sets may not be used again, and can be removed safely. MDE includes an interface to selectively evict elements from memory. The user must ensure that no reference to a given property set to be evicted remains within the program (in the form of indices) so that integrity is maintained. The implementation of peripheral features such as reference counting in such cases is left up to the user.

\paragraph*{Serialization.} MDE includes an interface to convert serialize the entire state of a construction of MDE nodes into a JSON file. A system like this may be especially useful in cases where analysis occurs in stages, and the stages do not share memory or happen in different runs. The memoization effort performed in earlier stages can be utilized in later stages to speed up the analysis.

\section{Complexities in the C++ Implementation of VASCO-LFCPA}
\label{sec:vasco-complexity}
Construction of the VASCO-LFCPA analysis on top of the LLVM framework is a non-trivial task and requires several important considerations to be made both in terms of the algorithm itself and LLVM's program representation. Here, we describe some of the complexities involved in implementing the analysis.

\paragraph*{Implementation of Two Separate Analyses.}
VASCO-LFCPA does not only contain an implementation of points-to analysis, but it contains an implementation of liveness analysis as well. As described earlier in~\Cref{sec:background:pta}, LFCPA interleaves liveness and points-to analyses. Liveness is a backward data-flow analysis, propagating information from exit to entry node, whereas points-to analysis is forward, propagating information from entry to exit node of a control flow graph. Thus, the possible statements in a given program have to be handled for both liveness analysis and points-to analysis, adding to the complexity of the implementation.

\paragraph*{Accommodating LLVM IR Features.}
Information regarding pointers in a program can be present in a significant number of instructions understood by LLVM's intermediate representation language~\cite{web-llvmlangref}, such as \texttt{load}, \texttt{store}, \texttt{phi}, and \texttt{GetElementPtr} instructions. All of these have to be handled individually on a case-by-case basis for computation of precise liveness and points-to information. Other features, such as scoping of local variables, mapping calling parameters and return statements to the formal definition of the function for analysis-specific reasons are also handled by the implementation.

\paragraph*{Resolution of Contexts and Indirect Calls.}
In an effort to achieve a precise interprocedural analysis and full context sensitivity, VASCO-LFCPA resolves indirect calls made with the use of function pointers through the points-to information reaching the point of the call. Since calling contexts in VASCO are distinguished based on incoming data-flow values at the call-site, a context may have to be re-analyzed in case of new information flowing to a point where an indirect call is made. VASCO has many other aspects that need to be handled such as management of the queuing of tasks within an analysis as well as determination of when analysis information reaches a fixed point in a context. Matching contexts of liveness analysis and points-to analysis is easy when call strings are used to represent them. However with value-sensitive contexts, the domains of the data-flow values are different and this matching requires novel ideas.

\section{Additional Motivations based on Past Analysis Work}
\label{sec:additional-motiv}
A significant amount of work in our research group at IIT Bombay involves data structures and mechanisms that we consider to be potential good use cases for MDE. Although the integration work for MDE onto the existing implementations and subsequent testing is yet to be completed, we believe that the integration will lead to good performance and memory efficiency gains. The following examples serve to strengthen our case for MDE.

\subsection{Heap Reference Analysis (HRA)}
Heap Reference Analysis~\cite{hra} is a program analysis that uses a storeless model of memory, and determines possible access paths of the members of an object allocated in the heap. For example, consider a linked list node with two elements, \texttt{data} and \texttt{next}, where \texttt{next} points to another linked list node allocated somewhere. We may have a statement like \texttt{node = node.next} within a loop at some point in the program. HRA builds a \textit{deterministic finite automaton} (DFA) that represents this pattern. This means that we are passing around a set of DFAs that represent these access paths across the control flow of the program.

We consider this to be a good use case for MDE because, like in points-to analysis, we are simply propagating graphs across the program, and there is potential for leveraging the redundancies and sparseness of data in the analysis with MDE. One additional augmentation we make in out in-progress implementation is that we make in this case is that we also store the incoming edges for a particular node as well. This allows us to replace subgraphs starting from the current node by backtracking to one of the previous nodes, which is a common operation in the analysis. The current structure can be notationally described as the following:
\[
\begin{aligned}
\nonumber\textit{HRAGraphMDE} &=
	\text{MDENode}(\textit{var}, \textit{normalop}, \textit{HRANodeMDE}, \textit{HRANodeMDE}) \\
\nonumber\textit{HRANodeMDE} &=
		\text{MDENode}(\textit{var}, \textit{normalop})
\end{aligned}
\]

\subsection{Generalized Points-to Graphs (GPG)}
\label{subsec:additional-motiv:gpg}
Generalized Points-to Graphs~\cite{gpg-analysis} represent points-to relationships with two integers denoting the amount of indirection between two variables instead of the concrete set of possible relationships. This summarizes the pointer assignment information within a given control flow graph, which then gets progressively simplified by an analysis to represent the points-to information as compactly as possible.

In the in-progress implementation at the time of writing, the edges are represented as a tuple of 4 values. These tuples are put into a set to denote a graph, and these graphs are put into another set to represent the summaries for control-flow blocks. This set-of-sets structure isn't directly modeled by MDE, but instead we model this implicitly like we discussed earlier in~\Cref{sec:gen-mde-cfg}. That is, make the parent MDE nodes store sets of indices valid within the child MDE nodes. While there are no nesting methods used here directly, we are still able to represent aggregate nested data structures. It can be notationally described as the following:
\[
\begin{aligned}
\nonumber\textit{GPSummaryMDE} &=
	\text{MDENode}(\textit{GPGraphMDE.ids}, \textit{normalop}) \\
\nonumber\textit{GPGraphMDE} &=
	\text{MDENode}(\textit{gptuple}, \textit{normalop}) \\
\end{aligned}
\]
Where \textit{GPGraphMDE.ids} is the domain of sets of valid set indices in \textit{GPGraphMDE}, and \textit{gptuple} is the domain of the set of the aforementioned tuples.

It may also be possible that we may change the representation structure entirely and use a representation similar to what we use for points-to sets. The different possible configurations have to be explored to see which exposes the maximum redundancy in data. Such configurations are currently under active exploration by us, and in the future we hope to obtain more results over the correlation of different constructions of MDE and their impact on efficiency.

\section{MDE Library Usage Examples}
This section contains various usage examples for the MDE C++ library. Some of these examples replicate scenarios presented in earlier sections.

\begin{lstlisting}[label={code:simpledemo}, language=c++,caption={Equivalent sample code for~\protect\Cref{subsec:mdedef:nonnestdemo}}]
#include <iostream>
#include <string>

// MDE definitions are placed in this header.
#include "mde/mde.hpp"

// This is the MDENode class definition.
//
// MDEConfig is a struct containing typedefs that act as parameters
// for the MDENode, such as the less-than, equality comparators,
// hash functions in the form of Functors (eg. std::less). To provide
// customizations, the typedefs can be overridden by deriving this
// struct.
//
// This defines an MDENode storing sets of strings. Strings are used
// for illustrative purposes.
using MDE = mde::MDENode<mde::MDEConfig<std::string>>;

// These are peripheral types we alias for convenience.
using Index = MDE::Index;
using Elem = MDE::PropertyElement;

// MDE object declaration.
MDE l;

int main() {

  // Strings we will use as elements.
  std::string a = "a", b = "b" , c = "c", d = "d";

  // Define two sets.
  Index i1 = l.register_set({a, b, c});
  Index i2 = l.register_set({a, b, d});

  // Get the union between them.
  Index i3 = l.set_union(i1, i2);

  std::cout
    << "Set {a, b, c} has index " << i1 << std::endl
    << "Set {a, b, d} has index " << i2 << std::endl
    << std::endl;

  std::cout << "Contents of Index " << i3 << ": ";

  // Print out all of the elements in i3.
  for (const Elem &elem : l.get_value(i3)) {
    std::cout << elem << " ";
  }
  std::cout << std::endl << std::endl;

  // This will dump all of the data present in the MDE.
  std::cout << l.dump();
}


/*
Output:

Set {a, b, c} has index 1
Set {a, b, d} has index 2

Contents of Index 3: a b c d

{
    Unions: (Count: 1)
      {(1,2) -> 3}

    Differences:(Count: 0)

    Intersections: (Count: 0)

    Subsets: (Count: 2)
      (2,3) -> sub
      (1,3) -> sub

    PropertySets: (Count: 4)
      0 : { }
      1 : { a b c }
      2 : { a b d }
      3 : { a b c d }
}
*/
\end{lstlisting}

\begin{lstlisting}[label={code:nesteddemo}, language=c++,caption={Equivalent sample code for~\protect\Cref{subsec:nestedcons:operdemo}}]
#include <iostream>
#include <ostream>
#include <string>

#include "mde/mde.hpp"

// Definition of child MDE.
using ChildMDE = mde::MDENode<mde::MDEConfig<std::string>>;

// Peripheral types we alias for convenience.
using ChildIndex = typename ChildMDE::Index;
using ChildElem = typename ChildMDE::PropertyElement;

// Definition of the parent MDE.
//
// Note the new parameter, NestingBase, which takes the key type of
// the parent MDE, and the class definition of the child MDE.
// NestingBase can take any number of MDE class parameters that are to
// be accounted as children of the parent.
using MDE =
  mde::MDENode<
    mde::MDEConfig<std::string>,
    mde::NestingBase<std::string, ChildMDE>>;

// Peripheral types we alias for convenience.
using Index = typename MDE::Index;
using Elem = typename MDE::PropertyElement;

// Child MDE object declaration.
ChildMDE cl;

// Parent MDE object declaration.
// It accepts a std::tuple containing references to the child MDEs.
// Its size and constituent types are determined at compile time.
MDE l({cl});

int main() {
  std::string a = "a", b = "b", c = "c", p = "p", q = "q", s = "s",
              t = "t", u = "u";

  // First we define sets in the child MDE to use in the parent MDE
  ChildIndex c1 = cl.register_set({p, q});
  ChildIndex c2 = cl.register_set({s, t});
  ChildIndex c3 = cl.register_set({t, u});

  // Now we define sets in the parent MDE. Note the child set index
  // present in another set of curly braces in each element of the
  // list. We require them because child indices are stored in a
  // tuple in each element.
  Index J = l.register_set({
    {a, {c1}},
    {b, {c2}},
  });

  Index K = l.register_set({
    {b, {c3}},
    {c, {c1}}
  });

  // Now we perform the union.
  Index result = l.set_union(J, K);

  std::cout << "Set J index: " << J << std::endl
            << "Set K index: " << K << std::endl
            << "Result set index: " << result << std::endl
            << "Result set contents: " << std::endl;

  for (const Elem &elem: l.get_value(result)) {
    // First we print out the entire element.
    std::cout << "  " << elem << " :: { ";

    // Then we traverse its children.
    // We are retrieving the Child MDE index from the value tuple.
    // value<0> returns the first index of the tuple.
    for (const ChildElem &childelem: cl.get_value(elem.value<0>())) {
      std::cout << childelem << " ";
    }
    std::cout << "}" << std::endl;
  }
  std::cout << std::endl;

  // Dump all of the data.
  std::cout << "@@@@ Pointee MDE @@@@" << std::endl;
  std::cout << cl.dump() << std::endl << std::endl;
  std::cout << "@@@@ Points-to MDE @@@@" << std::endl;
  std::cout << l.dump() << std::endl;

  return 0;
}

/*
Output:

Set J index: 1
Set K index: 2
Result set index: 3
Result set contents:
  a -> [ 1 ] :: { p q }
  b -> [ 4 ] :: { s t u }
  c -> [ 1 ] :: { p q }

@@@@ Pointee MDE @@@@
{
    Unions: (Count: 1)
      {(2,3) -> 4}

    Differences:(Count: 0)

    Intersections: (Count: 0)

    Subsets: (Count: 2)
      (3,4) -> sub
      (2,4) -> sub

    PropertySets: (Count: 5)
      0 : { }
      1 : { p q }
      2 : { s t }
      3 : { t u }
      4 : { s t u }
}


@@@@ Points-to MDE @@@@
{
    Unions: (Count: 1)
      {(1,2) -> 3}

    Differences:(Count: 0)

    Intersections: (Count: 0)

    Subsets: (Count: 2)
      (2,3) -> sub
      (1,3) -> sub

    PropertySets: (Count: 4)
      0 : { }
      1 : { a -> [ 1 ] b -> [ 2 ] }
      2 : { b -> [ 3 ] c -> [ 1 ] }
      3 : { a -> [ 1 ] b -> [ 4 ] c -> [ 1 ] }
}
*/
\end{lstlisting}

\begin{lstlisting}[label={code:2elemdemo}, language=c++,caption={Sample code depicting a two-child configuration in MDE.}]
#include <iostream>
#include <ostream>

#include "mde/mde.hpp"

// Definition of the first child.
using Child1MDE =
  mde::MDENode<mde::MDEConfig<double>>;

// Aliases.
using Child1Index = typename Child1MDE::Index;
using Child1Elem = typename Child1MDE::PropertyElement;

// Definition of the second child.
using Child2MDE =
  mde::MDENode<mde::MDEConfig<int>>;

// Aliases.
using Child2Index = typename Child2MDE::Index;
using Child2Elem = typename Child2MDE::PropertyElement;

// Definition of the parent MDE.
// Note the presence of Child1MDE and Child2MDE in the parameter list
// of NestingBase.
using MDE =
    mde::MDENode<
      mde::MDEConfig<int>,
      mde::NestingBase<int, Child1MDE, Child2MDE>>;

// Aliases.
using Index = typename MDE::Index;
using Elem = typename MDE::PropertyElement;

// Object definitons.
Child1MDE cl1;
Child2MDE cl2;
MDE l({cl1, cl2});

int main() {
  // Registering sets in Child 1
  Child1Index a1 = cl1.register_set_single(212.23);
  Child1Index b1 = cl1.register_set({22.2, 33.122, 33.4});

  // Registering sets in Child 2
  Child2Index a2 = cl2.register_set_single(212);
  Child2Index b2 = cl2.register_set({22, 33, 34});

  // Register a set in parent.
  Index c = l.register_set({
    {12, {a1, a2}},
    {14, {a1, b2}},
  });

  // Register a single-element set.
  Index f = l.register_set_single({12, {b1, b2}});

  // Perform set union.
  Index result = l.set_union(c, f);
  std::cout << "Result set index: " << result << std::endl
            << "Result set contents: {" << std::endl;

  // Iterate over both children of each element present in the result.
  for (const Elem &elem: l.get_value(result)) {
    std::cout << "    " << elem << " :: { " << std::endl
              << "        Child 1: ";

    // Access the first child with the index 0.
    for (const Child1Elem &j: cl1.get_value(elem.value<0>())) {
      std::cout << j << " ";
    }

    std::cout << std::endl
              << "        Child 2: ";

    // Access the first child with the index 1.
    for (const Child2Elem &j: cl2.get_value(elem.value<1>())) {
      std::cout << j << " ";
    }

    std::cout << std::endl;
    std::cout << "    }" << std::endl;
  }
  std::cout << "}" << std::endl << std::endl;

  // Dump all MDENodes.
  std::cout << "@@@@ Child 1 MDE @@@@" << std::endl;
  std::cout << cl1.dump() << std::endl << std::endl;
  std::cout << "@@@@ Child 2 MDE @@@@" << std::endl;
  std::cout << cl2.dump() << std::endl << std::endl;
  std::cout << "@@@@ Main MDE @@@@" << std::endl;
  std::cout << l.dump() << std::endl;

  return 0;
}

/*
Output:

Result set index: 3
Result set contents: {
    12 -> [ 3 3 ] :: {
        Child 1: 22.2 33.122 33.4 212.23
        Child 2: 22 33 34 212
    }
    14 -> [ 1 2 ] :: {
        Child 1: 212.23
        Child 2: 22 33 34
    }
}

@@@@ Child 1 MDE @@@@
{
    Unions: (Count: 1)
      {(1,2) -> 3}

    Differences:(Count: 0)

    Intersections: (Count: 0)

    Subsets: (Count: 2)
      (2,3) -> sub
      (1,3) -> sub

    PropertySets: (Count: 4)
      0 : { }
      1 : { 212.23 }
      2 : { 22.2 33.122 33.4 }
      3 : { 22.2 33.122 33.4 212.23 }
}


@@@@ Child 2 MDE @@@@
{
    Unions: (Count: 1)
      {(1,2) -> 3}

    Differences:(Count: 0)

    Intersections: (Count: 0)

    Subsets: (Count: 2)
      (2,3) -> sub
      (1,3) -> sub

    PropertySets: (Count: 4)
      0 : { }
      1 : { 212 }
      2 : { 22 33 34 }
      3 : { 22 33 34 212 }
}


@@@@ Main MDE @@@@
{
    Unions: (Count: 1)
      {(1,2) -> 3}

    Differences:(Count: 0)

    Intersections: (Count: 0)

    Subsets: (Count: 2)
      (2,3) -> sub
      (1,3) -> sub

    PropertySets: (Count: 4)
      0 : { }
      1 : { 12 -> [ 1 1 ] 14 -> [ 1 2 ] }
      2 : { 12 -> [ 2 2 ] }
      3 : { 12 -> [ 3 3 ] 14 -> [ 1 2 ] }
}
*/
\end{lstlisting}

\end{document}